# Optimizing Data Lakes' Queries

Thesis submitted in partial fulfillment
of the requirements for the degree of
"DOCTOR OF PHILOSOPHY"

by

# Gregory Weintraub

Submitted to the Senate of Ben-Gurion University
of the Negev

December 31, 2024

Beer-Sheva

# Optimizing Data Lakes' Queries

**Thesis submitted in partial fulfillment
of the requirements for the degree of
"DOCTOR OF PHILOSOPHY"**

by

**Gregory         Weintraub**

Submitted to the Senate of Ben-Gurion University
of the Negev

Approved by the advisors: Prof. Ehud Gudes and Prof. Shlomi Dolev

Approved by the Dean of the Kreitman School of Advanced Graduate Studies

December 31, 2024

Beer-Sheva

# Abstract

Cloud data lakes provide a modern solution for managing large volumes of data. The fundamental principle behind these systems is the separation of compute and storage layers. In this architecture, inexpensive cloud storage is utilized for data storage, while compute engines are employed to perform analytics on this data in an "on-demand" mode. However, to execute any calculations on the data, it must be transferred from the storage layer to the compute layer over the network for each query. This transfer can negatively impact calculation performance and requires significant network bandwidth.

In this thesis, we examine various strategies to enhance query performance within a cloud data lake architecture. We begin by formalizing the problem and proposing a straightforward yet robust theoretical framework that clearly outlines the associated trade-offs. Central to our framework is the concept of a "query coverage set," which is defined as the collection of files that need to be accessed from storage to fulfill a specific query. Our objective is to identify the minimal coverage set for each query and execute the query exclusively on this subset of files. This approach enables us to significantly improve query performance.

We study prior art to identify gaps in existing state-of-the-art solutions, and then we develop novel query optimization techniques for cloud data lake queries across three different domains:

1. **Indexing**: Indexing is a traditional method for improving query performance in relational databases. However, it is rarely applied in big data environments due to the enormous scale and management complexity. We have developed a novel indexing scheme specifically designed for cloud data lakes, based on the concept of a query coverage set. Our initial attempt focused on the "needle in a haystack" type of queries [1, 2]. We designed an index that maps column values to the coverage sets of the queries that seek those values. This index is stored in the data lake alongside the actual data and is created and updated using parallel algorithms. For experimental evaluation, we collaborated with a large enterprise that manages cloud data lakes at a petabyte scale. Subsequently, we broadened this scheme to accommodate a general scenario [3, 4]. In the general case, we defined an optimization problem that, if solved, can demonstrably accelerate data lake queries. We proved that this problem is NP-hard and proposed heuristic approaches to overcome its hardness. Our prototype implementation, along with extensive evaluations based on the TPC-H benchmark, showed a 30-fold improvement in query execution time.
2. **Caching**: Caching is another well-known method for enhancing query performance, typically involving the storage of intermediate query results for reuse in subsequent queries. However, in a big data environment, this approach can be problematic, as intermediate results may be particularly large, making traditional caching impractical. Our solution [5, 6] addresses this issue by caching metadata (coverage sets) instead of the actual data. This approach is based on concepts from both database and computational geometry fields. Experiments based on the TPC-H benchmark demonstrate its feasibility and efficiency.
3. **Genetic Data**: The two previous techniques are general-purpose and can be applied to any relational data. However, when we know in advance the type of data we are dealing with, we can design even more effective solutions. We collaborated with researchers from the Laboratory of Human Genetics at Ben-Gurion University to create GeniePool [7-9], a genomic repository built on a cloud data lake. By strategically organizing genetic data in the cloud, we can efficiently calculate the coverage set for any given query. The repository we established is widely used by genetic researchers and clinicians.

The primary research objective of this thesis is to discover new methods for query optimization in emerging yet crucial storage systems – cloud data lakes. We successfully developed innovative solutions across three distinct areas: indexing, caching, and genetic data. These solutions are grounded in a solid theoretical framework and have been validated through extensive experimental evaluations. Additionally, the solutions for indexing and genetic data have been applied in real-world scenarios.

Alongside the core focus of our research on query optimization techniques, we believe our following contributions may be of independent interest:

- **Theoretical model for data lake queries** – In [4], we mathematically formalized several key features of cloud data lakes, such as query coverage, tight coverage, tightness and coverage degrees, and the balanced query coverage problem. This model enables a comprehensive analysis of poor query performance in cloud data lakes and allows us to propose efficient solutions. We believe that this theoretical framework will be valuable to both researchers and engineers working with cloud data lakes.
- **Mapping of the query containment problem to geometric space** – In [6], we developed a method to improve the runtime of cache checks by mapping queries into multidimensional intervals. We then transformed the interval containment problem into a range search problem. This finding is likely to be of interest to the broader data management community.
- **GeniePool** – We created a real-world genetic application [7-9] that applies theoretical concepts from this research, providing a unique service to genetic researchers and clinicians in their efforts to address complex genetic disorders.

The results of this thesis have been published and presented at various venues [1-9]. Some of these publications include "work in progress." The most comprehensive and current works for each research topic are as follows:

- Indexing [4]
- Caching [6]
- Genetic Data [8, 9]

This work was carried out under the supervision of

Prof. Ehud Gudes and Prof. Shlomi Dolev

In the Department of Computer Science

Faculty of Natural Science

# Acknowledgments

There are many people without whom this thesis would not have been possible, and I would like to express my gratitude to them.

First, I want to thank my advisors, Professor Ehud Gudes and Professor Shlomi Dolev, for their constant support and guidance throughout this journey.

I am also thankful to the many people at IBM who helped me to pursue my dream of combining an industry career with academic research, including Nadav Katzenell, Edy Stein, Nati Braha, Yaron Wolfsthal, Gadi Katsovich, Lior Haber, and many others.

During my studies, I had the opportunity to collaborate with genetic researchers Professor Ohad Birk and PhD candidate Noam Hadar. It was a rewarding experience to step outside of my comfort zone and explore a completely new field. In this collaboration, Noam focused on the genetics aspect while I handled the algorithmic aspects and the innovations in the data lakes scope. Together, we developed a real-world application that is now used by many people worldwide. This project highlighted the importance of interdisciplinary research, where the whole is greater than the sum of its parts.

Finally, I want to express my deep gratitude to my family. I am thankful to my parents for instilling in me a curiosity and eagerness for knowledge, my beloved wife, Sasha, for her endless patience, and my daughter, Dina, for her unwavering belief in my success.

# Contents

# List of Figures

# List of Tables

# 1  Introduction

Traditionally, storage systems have been favoring data locality (meaning they wanted to be as close to the data as possible to speed up calculations on the data). In single-node databases, data locality occurs trivially, whereas in shared-nothing distributed systems [10-13], data locality is achieved by performing computation on the same machines that store the data.

However, with the rise of cloud technologies, a new family of storage systems has emerged – cloud object stores [14-16]. These systems provide object storage service [17] through a web interface. Users create buckets, and each bucket may contain multiple binary objects uniquely identified within the bucket by a string key.

Cloud object stores are often recognized for their cost-effectiveness [18, 19, 20, 21, 94]. As a result, they are heavily used as the main building block of enterprise data repositories that have come to be known as *cloud data lakes* [18, 19, 95, 96].

The main feature of the cloud data lakes is that they store data in cloud object stores and, as a result, do not follow the traditional shared-nothing architecture but, instead, disaggregate the compute layer from the storage layer. A typical approach for storing enterprise relational data in a cloud data lake looks as in Fig. 1:

1. Raw events arriving from the outside of the organization (e.g., from sensors) are processed by the dedicated Extract-Transform-Load (ETL) job [23, 24] and uploaded to the object store in a columnar format [25, 26]. The columnar format provides efficient column-level compression and speeds up projection queries.
2. Data lake files are usually organized into partitions [27] according to the domain business logic, which allows skipping irrelevant files during the reads (e.g., "year" and "month" in Fig. 1).
3. Query engines [27-29] running on dedicated clusters execute SQL queries over the data in the lake.

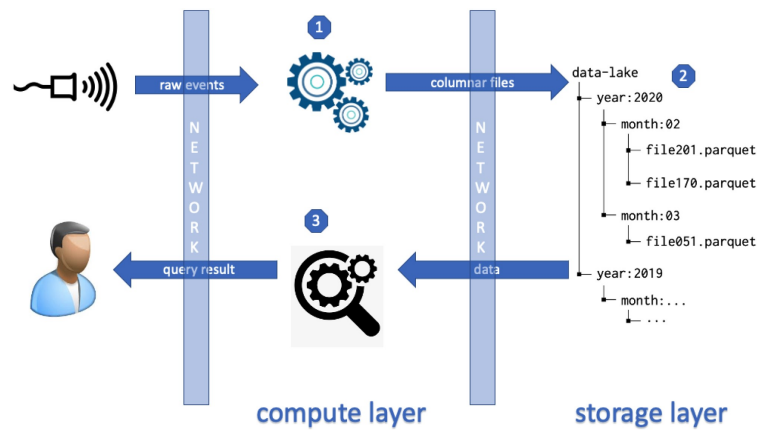

*Figure 1 Data Lake System Model*

Cloud data lakes introduce both benefits and challenges and, not surprisingly, are identified as a promising research direction by recent studies [19, 22]. Main benefits include:

- *Independent scaling of compute and storage layers*: For example, users can have very large data lakes (petabytes and beyond) but query them occasionally with an "on-demand" cluster. Such a scenario makes perfect sense for analytical use cases, but it is supported neither in traditional databases nor in Hadoop [12], where hardware is optimized for both storage and compute.
- *No vendor lock-in*: Disaggregated architecture and usage of standard tools in both storage and compute layers result in users' flexibility in moving between different cloud providers. This also simplifies the adoption of new technologies, which in turn promotes research and innovation in this area.
- *General cloud benefits*: The cloud model brings a lot of advantages: pay-as-you-go, expert management, economies of scale resulting in a lower operational cost, and more [21].

The drawbacks are:

- *Poor query performance*: Data should be moved from the storage layer to the compute layer for each calculation, and that significantly hurts query performance. This issue is particularly problematic in interactive queries, as users have to wait an excessive amount of time for their results.
- *Difficulties with handing updates*: Object stores support only simple "put" operations, and as a result, there is no atomicity across multiple objects. In addition, mutations inside objects are not supported, so when there is a need to change the content of some file, the entire file should be rewritten.
- *Security issues*: There are many security-related challenges in the cloud model, including privacy, integrity, and availability issues [30].

In this thesis, we focus on the first drawback (poor query performance); we discuss and define it formally in the following section.

The main contributions of this thesis can be summarized as follows:

1. *Development of a Theoretical Framework:* This framework formally defines the problem of query performance in cloud data lakes (Chapter 2).
2. *Critical Review of Related Work:* A thorough examination of existing literature on query execution performance in cloud data lakes was conducted, identifying their limitations (Chapter 3).
3. *Novel Indexing Scheme:* We developed a new indexing scheme for cloud data lakes (Chapter 4). This scheme includes a definition of the optimization problem that can provably improve the speed of data lake queries. We proved that the problem is NP-hard and proposed heuristic approaches to address its complexity.
4. ***Novel Caching Scheme*:** We introduced a novel caching scheme for cloud data lakes (Chapter 5). This scheme maps the data management problem to a geometric space, providing an efficient solution.
5. ***Real-World Application for Genetic Data in the Cloud:*** Our theoretical concepts were applied in a real-world context within the genetics domain. We collaborated with genetic researchers to create a repository of human mutations, stored in a cloud data lake and queried using our optimization scheme tailored specifically for their needs (Chapter 6).

The results of this thesis have been published and presented at various venues. Below is a brief overview:

1. In papers [1] and [2], we introduced the initial concepts for applying indexing in cloud data lakes, focusing primarily on "needle in a haystack" type of queries.
2. In papers [3] and [4], we expanded on the approach from [1] and [2] to support a more general scenario.
3. In [5], we introduced initial ideas for our caching scheme, which was significantly enhanced in [6] through the mapping to a geometric space.
4. Our preliminary concepts on the application of cloud data lakes in the genetics domain were discussed in [7]. The comprehensive findings on this topic were published in [8], where we unveiled the first version of our application, GeniePool. In [9], we presented an extended version of the system, GeniePool 2.0, which was developed based on user feedback.

The rest of the thesis is structured as follows:

- In Chapter 2, we formally define the problem and research goals.
- In Chapter 3, we review related work.
- In Chapters 4, 5, and 6, we present our solution to the problem in different domains – indexing, caching, and genetic data, respectively.
- We conclude in Chapter 7.

# 2   Problem Statement and Preliminaries

Let us introduce the problem first using a simple example (we are going to use this example throughout the thesis). Consider a typical metric data presented in Table 1. Let us assume that this table is stored in the cloud data lake depicted in Fig. 1, such that records 1-3 are stored in "file201", records 4-6 are stored in "file170", and records 7-9 in "file051". Files' format can be any of the standard supported formats (e.g., Parquet, ORC, CSV, JSON, etc.).

Table 1 Sample Metric Data

| rec-id | date | metric | val | ... | |
|---|---|---|---|---|---|
| 1 | 2020-02-10 | cpu | 47 | ... | file201 |
| 2 | 2020-02-14 | cpu | 58 | ... | |
| 3 | 2020-02-18 | memory | 11 | ... | |
| 4 | 2020-02-16 | memory | 8 | ... | file170 |
| 5 | 2020-02-20 | cpu | 88 | ... | |
| 6 | 2020-02-21 | cpu | 66 | ... | |
| 7 | 2020-03-13 | memory | 6 | ... | file051 |
| 8 | 2020-03-22 | cpu | 92 | ... | |
| 9 | 2020-03-28 | cpu | 71 | ... | |
| ... | ... | ... | ... | ... | ... |

Let us briefly review how the state-of-the-art query engines [27-29] would execute a typical query on Table 1 stored in the cloud data lake. For example, when a query engine receives the following query (Query 1), it reads the files from the storage (usually in parallel), scans them in-memory to find the records satisfying the predicate, and returns the result.

```sql
SELECT *
FROM   metrics_table
WHERE  metric = 'memory' AND val > 10
```

*Query 1*

In our example, only record three from file201 will be returned. However, all the data lake files should be read from the storage and processed, and since production data lakes might contain billions of files [31], this approach is extremely wasteful.

Even though this example might look oversimplified, many real-world scenarios have the same limitation. In fact, any query that needs only a small part of the data lake for its calculation will suffer from the same problem. Examples of such scenarios include:

- Looking for a specific record(s) in the data lake for troubleshooting.
- Fetching records related to a particular pattern for further analysis or ML training.
- Managing the General Data Protection Regulation (GDPR) compliance [32], where records related to a specific user should be found in the data lake upon user request.

So, intuitively, we would like to read only the "relevant" files for a given query (file201 in our example) and skip all the rest. Let us define the problem formally now.

## 2.1 Formal Problem Definition

The notation used throughout the thesis is presented in Table 2.

We model data lake tables according to the standard relational model [33]. Given a set of $m$ domains $D = \{D_1, D_2, ..., D_m\}$, a table (relation) $T^1$ is defined as a subset of the Cartesian product $D_1 \times D_2 \times ... \times D_m$. Each domain $D_i \in D$ has an associated column name $c_i \in L$, where $L$ is the set of all column names. $T$ is a set of tuples $\{t_1, t_2, ..., t_{|T|}\}$, where each tuple $t$ is a set of pairs $\{(c_i: v) \mid c_i \in L, v \in D_i, i \in \{1, ..., m\}\}$.

$T$ is stored in a cloud object store as a collection of files denoted by $F = \{f_1, f_2, ..., f_{|F|}\}$. $F$ is a partition of $T$, meaning that $\forall\, i \neq j : f_i \subseteq T, \bigcup F = T, f_i \cap f_j = \emptyset$. We assume each data lake file has an associated creation timestamp, and $F_{>ts}$ denotes files with a creation time older than $ts$. For the purpose of the formal definition, we assume that all the files are of the same size, so all the read operations from the cloud are equivalent from the cost perspective.

*Cost Model:* In our cost model, we assume that the cost of query engine operations is dominated by the reads from the remote storage and denote the cost of operation $x$ as $C(x)$ (i.e., to execute $x$, we need to perform $C(x)$ reads from the cloud storage). Our decision to use reads from the storage as the dominant component of the cost is consistent with the traditional DBMS approach [34], where the main cost metric is the number of I/O requests.

---

[1] We assume here a single relational table. The case of multiple tables and joins between them is briefly discussed in Section 4.5 and Section 5.2.4.

*Table 2 Notation*

| notation | description |
|---|---|
| $T$ | a data lake table |
| $L$ | column names |
| $D$ | domains |
| $m$ | number of columns in a data lake table |
| $F$ | data lake files |
| $F_{>ts}$ | data lake files with creation time older than *ts* |
| $Q$ | a data lake query |
| $F(Q)$ | files read by query Q from the storage |
| $P_Q$ | a predicate in the "where" condition of query Q |
| $S(P_Q, t)$ | tuple t satisfies predicate P of query Q |
| $Cov(X, Q)$ | query Q is covered by files X |
| $TC(Q)$ | tight coverage set of Q |
| $TD(X, Q)$ | tightness degree if the coverage set X of query Q |
| $CD(Q)$ | coverage degree of the query Q |
| $|x|$ | size of object x |
| $C(x)$ | cost of object x |
| $C_e(x)$ | estimated cost of object x |
| $I_c$ | data lake index on column c |
| $K$ | upper limit of the balanced coverage plan cost |
| $N$ | natural numbers $\{0, 1, 2, ...\}$ |
| $c.min, c.max$ | min and max values of column c |

***Definition 1*** *(data lake query):* We define a data lake query $Q$ as a standard SQL query on table $T$ and denote the predicate in the where clause of $Q$ as $P_Q$. We assume that $P_Q$ is given in a conjunctive normal form (CNF), which looks as follows: $(T_{11} \lor T_{12} \lor ...) \land (T_{21} \lor T_{22} \lor ...) ... \land (T_{n1} \lor T_{n2} \lor ...)$. CNF is a conjunction of *clauses* where a clause is a disjunction of *terms*. A term is a condition of type <column **op** value> (e.g., age > 40) or <column1 **op** column2> (e.g., salary < department-avg-salary), where columns are taken from $L$, values from domains in $D$, and $op \in \{=, \neq, \geq, \leq, >, <\}$.

If tuple $t$ from the file $f \in F$ satisfies $P_Q$ we denote it by $S(P_Q, t)$. Files that are read by query $Q$ from the storage are denoted by $F(Q) \subseteq F$

**Definition 2** *(query coverage):* Given a data lake query $Q$, $X \subseteq F$ covers $Q \leftrightarrow \forall f \in F \setminus X, \neg \exists t \in f, S(P_Q, t)$.

When $X$ satisfies Definition 2 for some data lake query $Q$, we say that $X$ covers $Q$ (meaning that $X$ contains all the files needed to satisfy $Q$) and denote it by *Cov (X, Q)*. We call such $X$ a *coverage set* of $Q$. Note that for any $Q$, holds *Cov (F, Q)*.

**Definition 3** *(query tight coverage):* Given a data lake query $Q$, $X \subseteq F$ tightly covers $Q \leftrightarrow Cov(X, Q) \wedge \neg \exists f \in X, \forall t \in f, \neg S(P_Q, t)$.

When $X$ satisfies Definition 3 for some data lake query $Q$, we say that $X$ tightly covers $Q$ (meaning that $X$ contains all the files needed to satisfy $Q$ and only them). We call such $X$ a tight coverage set of $Q$ and denote it by *TC (Q)*. By considering the example query above (Query 1), we can say that the query is *covered* by {"file201", "file170", "file051"} but not by {"file170", "file051"} and is *tightly covered* only by {"file201"}.

We can also extend the definitions of *coverage* and *tight coverage* to the record level (instead of the files level defined in Definitions 2, 3). In this case, tight coverage of the query $Q$ in a record level is simply the actual record IDs returned by $Q$, and a coverage set of $Q$ in a record level is any subset of table record IDs that contains the result record ids of $Q$.

If, for any data lake query $Q$, we could (efficiently) compute $X$ such that $X = TC(Q)$ we would be able to significantly improve query performance in a cloud data lake architecture by accessing only files in $X$ instead of all those in $F$ (and in most real-world scenarios $|X| \ll |F|$). In fact, as will be shown below, in many cases, finding the exact tight coverage might be too complicated, and we can be content with some coverage set that is not tight but still can help us improve query performance. For such scenarios, the definition of *tightness* and *coverage* degrees might be useful.

**Definition 4** *(tightness degree):* Given a data lake query $Q$, for any coverage set $X$ of $Q$, the tightness degree of $X$ is defined as:

$$TD(X, Q) = \begin{cases} if\ |TC(Q)| < |F|, & 1 - \dfrac{|X| - |TC(Q)|}{|F| - |TC(Q)|} \\ otherwise, & 0 \end{cases} \quad (1)$$

Intuitively, the tightness degree shows to what extent the given coverage set is close to the tight coverage set (1 means a perfect match).

***Definition 5*** *(coverage degree):* Given the data lake query $Q$, the coverage degree of $Q$ is defined as:

$$CD(Q) = \frac{|TC(Q)|}{|F|} \qquad (2)$$

Coverage degree shows what part of the data lake must be scanned by $Q$. The best coverage degree is 0 (when no files need to be accessed), and the worst is 1 (when all files are required).

Based on the above semantics, we can formulate the main research questions of this thesis as follows:

1. Can we develop an algorithm that, for any data lake query $Q$, can find a coverage set of $Q$, $X$, such that:
    a. tightness degree of $X$ is maximized
    b. cost of computing $X$ is minimized
    c. as a result of the above, the total execution time of $Q$ is reduced as much as possible
2. What are the practical considerations of applying the algorithm from question 1 in real-world systems?

Research question 1 above can be defined formally as the following multi-objective optimization problem [35]:

$$minimize\ \{C(X), 1 - TD(X,Q)\} \qquad (3)$$

$$subject\ to\ X \subseteq F, Cov(X,Q)$$

The above is the formal definition of the general problem. In Chapters 4, 5, and 6, we present definitions and solutions for some sub-problems of it.

# 3 Related Work

In this chapter, we discuss relevant previous work. In the following chapters (Chapter 4, Chapter 5, and Chapter 6), we will explore additional significant works that relate specifically to each chapter's focus.

As previously mentioned, a key feature of cloud data lake architecture is the separation of the compute and storage layers (see Fig. 1 and Fig. 2). In the sections below, we provide an overview of each of these layers. Section 3.1 reviews the storage layer, Section 3.2 examines the compute layer, and Section 3.3 explores prior research related to query optimization in cloud data lakes.

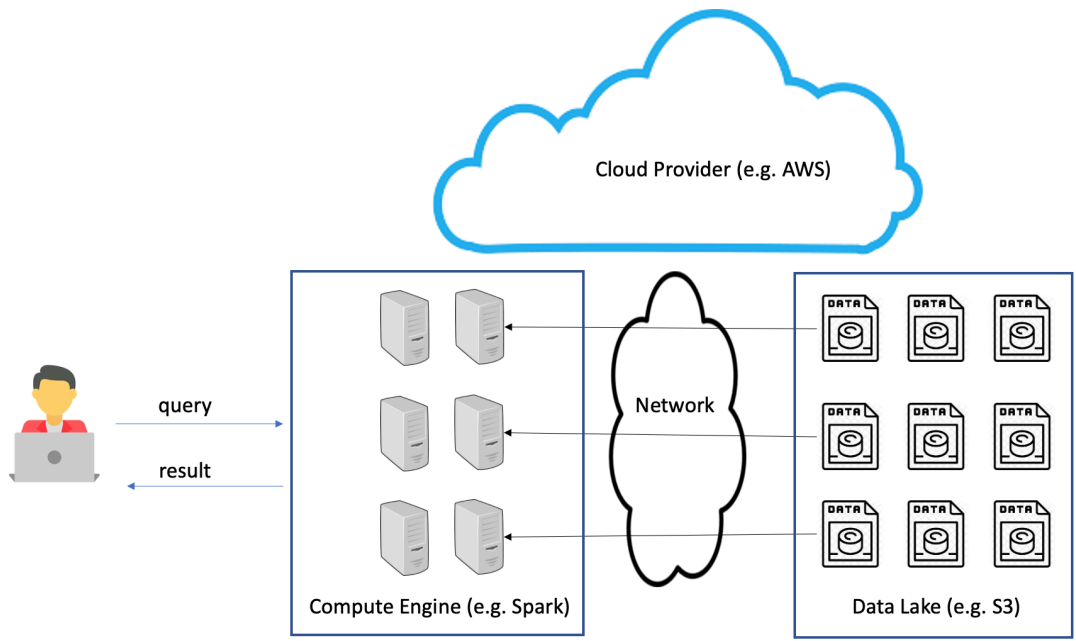

*Figure 2 Cloud Data Lake Architecture*

## 3.1 Cloud Data Lakes

Since the 1980s, relational databases have dominated the data management industry [86]. However, in the 2000s, large tech companies such as Google and Amazon started experiencing difficulties with the traditional way of storage [10, 87].

In 2003, Ghemawat et al. introduced the Google File System (GFS) in their groundbreaking paper [10], marking a pivotal moment in the data management industry. GFS is a distributed file system built on several important observations that are still relevant today:
1. Component failures are the norm rather than exception.
2. Files are huge by traditional standards.
3. Once written, the files are only read, and often only sequentially.

The high-level architecture of GFS is presented in Fig. 3. A GFS cluster consists of a single *master* node and multiple *chunkservers*. Files are divided into fixed-size *chunks* and chunkservers store them on local disks. To ensure reliability, each chunk is replicated on multiple chunkservers, with a default replication factor of three. The master node maintains all system metadata, such as the mapping between files and chunks, as well as the locations of the chunks. Clients interact with the master node for metadata operations, while all data communication occurs directly with the chunkservers.

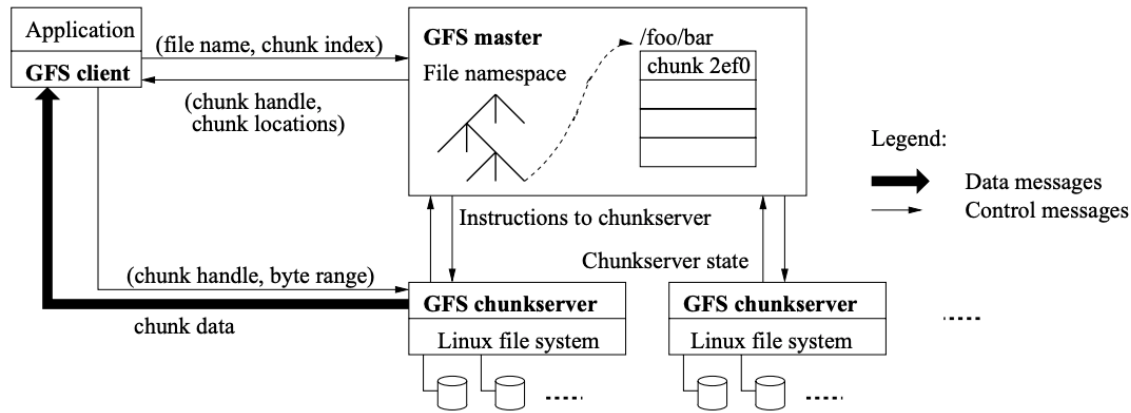

*Figure 3 GFS Architecture (taken from [87])*

Following the publication of the GFS paper, Yahoo began developing an open-source implementation of GFS, which later became known as the Hadoop Distributed File System (HDFS) [12]. HDFS was donated to the Apache Software Foundation as part of the Hadoop project in 2006.

HDFS adopts the general architecture of GFS but uses different terminology for its components. For instance, the master node is referred to as the NameNode, the chunkservers as DataNodes, and the chunks as blocks. HDFS gained significant popularity in the data management community, becoming a primary building block for numerous projects in both industry and academia [27, 88, 89, 90]).

In 2010, James Dixon, the CTO of Pentaho, coined the term "data lake." He described it as a concept where "*the contents of the data lake stream in from a source to fill the lake, and various users can come to examine, dive in, or take samples*" [91]. Unlike traditional data warehouses, which involve analyzing, filtering, transforming, and storing the data for future business needs, the data lake approach recognizes that "*there are many unknown questions about the data that will arise in the future.*" Therefore, it advocates storing the original raw data and allowing users to decide how and when to use it.

HDFS was an obvious choice for storing data lakes due to its high scalability and cost-effectiveness. Data was typically stored in open file formats like Apache Parquet [25] and Apache ORC [26], enabling various engines to query the data lake [23, 27, 28, 29, 89, 92].

The main idea of the columnar format is that instead of storing the data row by row, as is customary in relational databases, the data is stored column by column. The key advantages of column-wise storage include fast projection queries, efficient compression and encoding for specific columns, and effective data skipping, which will be discussed further in [Section 3.3.2](#).

Simultaneously with HDFS's rise, cloud object stores [14-16] emerged. Amazon's Simple Storage Service (S3), launched in 2006 [93], was the first of its kind and established the basic API for cloud object stores that later other vendors adopted. Unlike HDFS, which functions as a true file system, cloud object stores resemble key-value databases and only support simple *put (key, object)* and *get (key)* operations within a namespace known as a "bucket." Buckets must be created before use, and users have the ability to list objects inside their buckets.

It is common to construct object keys in a "file system" style. For example, we can simulate folder hierarchy by creating objects with keys "root/folder1/obj1" and "root/folder1/obj2". Object stores support metadata "List-Objects" operation that allows users to get the keys of all objects from the particular "folder" (e.g., list on "root/folder1/" prefix would return "obj1" and "obj2").

By treating every file as an individual object, cloud object stores can simulate file system behavior. And because they outperform HDFS in many important dimensions (scalability, durability, availability, and cost) [94], since 2015 they started replacing HDFS as the primary storage layer behind data lakes [19]. This approach of data lakes implemented on top of cloud object stores has come to be known as cloud data lakes [18, 19, 95, 96]. Nowadays, cloud data lakes are reported to be in use at "virtually all Fortune 500 enterprises" [19] and are expected to dominate as the "OLAP DBMS archetype for the next ten years" [96].

In summary, the key concept behind data lakes is to store vast amounts of data in its original (or minimally processed) format. This approach eliminates the need for scurpolous schema design and high development and maintenance costs typically associated with traditional data warehouses. Instead, data is streamed into a distributed storage system, making it immediately available for analytics. While the first generation of data lakes relied on HDFS for storage, recent years have seen cloud object stores establish themselves as the primary storage method, leading to the popular term "cloud data lakes."

## 3.2 Query Engines

As noted in the previous section, a significant turning point in the data management community was the publication of Google's GFS paper in 2003 [10]. However, this paper focused only on the storage aspect. In 2004, Google followed up with another pivotal paper titled "MapReduce: Simplified Data Processing on Large Clusters." [24].

MapReduce was designed to execute computations on vast amounts of data stored in GFS. Its main objective was to provide a framework that abstracts away low-level details such as parallelization, fault tolerance, data distribution, and load balancing, enabling developers to concentrate solely on business logic. This logic is defined functionally through *map* and *reduce* functions.

Together with GFS, MapReduce attracted significant attention. The Hadoop project, initiated by Yahoo and later donated to the Apache Software Foundation, consisted of two key components: HDFS (an implementation of GFS) and Hadoop MapReduce (an implementation of MapReduce). The Hadoop library quickly became synonymous with big data storage and computation, leading to the development of countless applications based on it [98, 99].

However, as the data lake approach emerged, the MapReduce paradigm fell short of meeting enterprise needs. Many enterprise users preferred querying their data lakes using SQL rather than MapReduce, which often requires extensive programming skills. Consequently, a variety of new systems, often known as *query engines* [97], were developed to enable SQL queries on top of data lakes.

Hive [27], originally developed at Facebook, was the first system of this kind. In Hive, SQL-like queries are compiled into a series of MapReduce jobs, which are executed within the MapReduce framework. A similar approach is employed in Spark SQL [28], where SQL queries run on top of the Apache Spark engine [23].

Another group of query engines does not rely on general-purpose frameworks like MapReduce or Spark but uses proprietary engines based on principles from parallel databases. The first such system was Google's Dremel [100], which inspired many other query engines, including well-known examples like Drill [101], Presto [29], and Impala [89].

All query engines share two main features:
1. *In-situ processing*: This means data is accessed in place – either in HDFS or a cloud object store – eliminating the need to load data into a database management system. This feature allows multiple query engines to query the same data lake simultaneously without issues
2. *Parallel processing*: Data is accessed in parallel, typically using a cluster of dedicated machines.

As data lakes transitioned to the cloud, query engines gradually included support for cloud data lakes. To facilitate this, cloud providers implemented an HDFS interface, allowing query engines to seamlessly operate on data lakes stored in both HDFS and cloud object stores.

In addition to query engines originally developed for Hadoop, new cloud-native engines emerged, such as AWS Athena [102] (based on Presto [29]) and Google BigQuery [47, 103] (based on Dremel [100]).

One important distinction between query engines that operate on data lakes stored in HDFS and those functioning on cloud data lakes is that, in HDFS, it is common to run computation processes close to the data. This means that the same machine is used for both storing the data and executing computations. In contrast, query execution in cloud data lakes is always completely separated from storage (Fig. 1, Fig. 2). This fundamental difference leads to a performance penalty for query execution in cloud data lakes, which is the primary focus of this thesis. In the following section, we will review existing approaches to address this issue.

## 3.3 Query Optimization in Cloud Data Lakes

The simplest method for executing queries in cloud data lakes is to read all the files. Clearly, *F covers* any *Q*. However, while this approach ensures that all queries are covered, it is not efficient, particularly when the query coverage is not *tight* – except in specific cases where *CD(Q)* equals 1. As a result, query performance tends to be poor. In the following sections, we will explore various techniques to enhance query performance in cloud data lakes.

### 3.3.1 Partitioning

One of the first suggested optimizations was data partitioning [27]. Consider again sample data in Table 1 where the table is partitioned by "year" and "month" columns. For queries whose predicates are based on partition columns, we can easily calculate the tight coverage set. For example, for the following query (Query 2), we need to access only files in folder "year=2020/month=02" – {"file201", "file170"}.

```sql
SELECT  *
FROM    metrics_table
WHERE   year = '2020' AND month = '02'
```
*Query 2*

Partitioning is a simple and powerful technique and is supported by all modern query engines (e.g., [13, 18, 23, 27, 29, 36]). Unfortunately, only a limited subset of table columns can be used in partitioning, while production tables may contain tens of thousands of columns [37]. As a result, global (cross-partition) queries cannot benefit from partitioning and need to scan all the data lake files.

Another limitation of the partitioning approach is that it cannot be applied to columns with high cardinality, as this would lead to the well-known "small files" problem [53].

### 3.3.2 Data Skipping

Another well-known approach [38] is to attach metadata to each data lake file and use it during the reads to skip irrelevant files. For example, the following query (Query 3) on the data in Table 1 can utilize a data skipping technique in the following way:

1. Each file contains a metadata section that stores, among other things, the minimum and maximum values for each column. In the case of Table 1, for the column "val," the minimum and maximum values per file are as follows: file201 = {min: 11, max: 58}, file170 = {min: 8, max: 88}, and file051 = {min: 6, max: 92}.
2. The query engine reads only the metadata section of each file, which is significantly smaller than the entire file, and evaluates whether the file can be skipped based on the predicate and the min/max values.
3. For Query 3, the metadata indicates that only file051 may contain relevant records (those with val > 90), allowing us to safely skip all other files.

```sql
SELECT *
FROM   metrics_table
WHERE  val > 90
```

*Query 3*

Columnar formats support metadata-based skipping out-of-the-box [25, 26, 39, 128] by storing the metadata and the data in the same file and relying on the fact that cloud object stores support reading of the particular sections of the file. To reduce the overhead of reading metadata from each file, recent studies suggest keeping all the metadata in a centralized place [18, 27, 37, 40] instead of per file.

Unfortunately, metadata-based skipping is very sensitive to data distribution and helps only in cases where the data is nicely clustered. For example, Query 1 above cannot skip any of the files based on metadata, while its tight coverage set contains only a single file ("file201").

There have been several approaches to optimize data organization during the writes so it will maximize the benefits of metadata-based skipping [18, 41, 42, 43]. For instance, in [18], the table can be reorganized using Z-Order [129] based on a user-defined set of attributes. This ensures that related information is stored within the same set of files. This method is computationally straightforward and creates locality across all specified dimensions. Once the data is clustered in this way, the data-skipping technique mentioned earlier becomes significantly more efficient.

### 3.3.3 Predicate Pushdown

Some cloud providers, in some cases (e.g., [44, 45]) support pushing the query predicate to the storage layer, so irrelevant records might be filtered out during the read operation, and only the relevant records would be returned to the compute layer. This technique is a great optimization, as we do not need to move huge amounts of data between storage and compute layers. However, we still perform a lot of time-consuming filtering operations; the only difference is that the operation is performed in a different place. In addition, the predicate pushdown operation can be charged a higher price by the cloud vendor than the regular read operation. The relation between pushdown operations and their cost was studied in PushdownDB [104].

The predicate pushdown technique does not help with skipping the reading of irrelevant files but makes this operation faster. Thus, it cannot help with the problem we consider in this thesis but can be combined with our techniques as an optimization.

### 3.3.4 Data Warehouses

Data warehouses have been used for decades to optimize query performance and support business intelligence (BI) inside enterprises. In the cloud era, modern data warehouses (such as Redshift [13] and Snowflake [46]) are used as a second tier in the data platform architecture: Data first is uploaded into the data lake (as described above) and later ingested into the data warehouse.

While certainly improving query performance, the data warehouse approach has the following significant drawbacks:

- *Maintenance complexity* - keeping two very large storage systems (data lake and warehouse) in sync is not a trivial engineering task, which requires building and maintaining complex ETL processes. In addition, the data warehouse will always have stale data compared to that of the data lake, which can cause issues with BI flows.
- *Cost* - in addition to the ETL cost, the same data is stored twice, hence the double storage cost.

Ideally, we would like to find a solution that adds data warehouse capabilities to the data lake instead of keeping two different storage systems for the same data. This novel approach has been named *lakehouse* in recent studies [19, 31, 48, 49]. Our solutions developed in this thesis are a step towards achieving the "lakehouse" horizon.

### 3.3.5 Table Format

Table format (e.g., Delta Lake [18], Apache Iceberg [50], and Apache Hudi [51]) is a novel approach to adding missing capabilities (e.g., transactions, schema evolution, query optimization) to the data lake architecture. The main idea is to add an additional layer of metadata between the files in the storage and the compute layer. This metadata can then be used, for example, to store information about schema changes, data mutation, and various statistics to improve query performance.

While table format is an important step towards the lakehouse vision (and in some works, it is already considered to be a lakehouse [31, 96], it still does not solve the main problem considered in this thesis: reading of (a lot of) irrelevant files from the storage. The main reason is the same as in data skipping ([Section 3.3.2](#)) - if we have a predicate that is based on columns that are not clustered, we are unable to fully utilize file statistics and may end up reading unnecessary files; our approaches tackle this problem effectively.

# 4 Indexing

## *Publications*

1. *Weintraub, G., Gudes, E. and Dolev, S., 2021, January. Needle in a haystack queries in cloud data lakes. In EDBT/ICDT Workshops (Vol. 93, p. 125).*
2. *Weintraub, G., Gudes, E. and Dolev, S., 2021, June. Indexing cloud data lakes within the lakes. In Proceedings of the 14th ACM International Conference on Systems and Storage (pp. 1-1).*
3. *Weintraub, G., 2023, August. Optimizing cloud data lakes queries. In Conference on Very Large Data Bases (VLDB 2023).*
4. *Weintraub, G., Gudes, E., Dolev, S. and Ullman, J.D., 2023. Optimizing cloud data lake queries with a balanced coverage plan. IEEE Transactions on Cloud Computing, 12(1), pp.84-99.*

## 4.1 Overview and Intuition

Let us first show what the problem is with the existing (naive) approach. Algorithm 1 presents the naive way of calculating the tight coverage set for a given query. As the algorithm reads all the data lake files from the cloud, its cost is $|F|$. Interestingly, that is also the best cost we can have in a general (worst-case) scenario, as assuming we can find the tight coverage set of an arbitrary query in sub-linear cost $o(F)$, would imply we can build a data structure of sub-linear size that can answer arbitrary search queries, and that would contradict the information-theoretic lower bound [54].

**Algorithm 1:** Naive Get Tight Coverage.
**Input:** data lake query $Q$, data lake files $F$
**Output:** $TC(Q)$

1: $result \leftarrow \emptyset$
2: **for** each file $f$ in $F$ **do**
3:     read $f$ from the cloud
4:     **for** each tuple $t$ in $f$ **do**
5:       **if** $S(P_Q, t)$ **then**
6:         $result \leftarrow result \cup f$
7:         break
8:       **end if**
9:     **end for**
10: **end for**
11: **return** $result$

*Algorithm 1*

Fortunately, for many practical scenarios, we can do better (in terms of the query performance cost) if we allow reducing the tightness degree. The main idea is, instead of looking for the *tight* coverage set of the given query, to find *some* coverage set that will result in an optimal *total* query execution time; we are looking for the coverage set $X$ such that the sum of the cost of finding $X$ and its size is minimized.

In our approach, for each data lake query, we focus on the "where" condition and look at each predicate clause separately ($C$ values in Fig. 4). There may be many coverage sets associated with each clause, and there may be many ways to compute each of these coverage sets ($P$ values in Fig. 4). We will assume that for each possible *coverage execution plan*, we can estimate (e.g., via statistics, caching, ML models, etc.) what is the expected cost and expected result of each plan (our approach for estimation is presented in Section 4.2.3). One example of a coverage plan is trivially to return $F$, which is a coverage set of any query. In this case, the cost of the plan is 0, as we do not read files from the cloud storage at all, and its result is $F$. Another example of a coverage plan is to run Algorithm 1. In this case, the cost is $|F|$, as we scan all the data lake files, and the result is $TC(Q)$, which can be any subset of $F$, depending on the query.

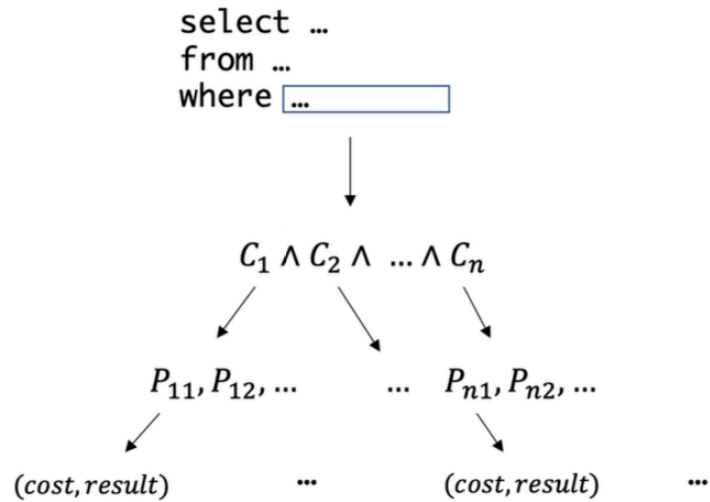

Figure 4 Query estimations structure

An important observation (defined formally in Theorem 1 below) is that the intersection of coverage sets of any subset of the query clauses is a coverage set of the original query. Based on this observation, we can explain the main idea of our approach (defined formally in Definition 6 below) as follows:

1. Given a query and its estimated values as in Fig. 4, we want to find a subset of clauses (and their corresponding coverage plans) such that the sum of their estimated costs and the size of the intersection of the coverage sets is minimized.
2. Then, we execute each of the coverage plans (in parallel), intersect their results and execute the original query on the files in the intersection only (which is a coverage set of the original query according to Theorem 1).

**Theorem 1:** Let $Q$ be a data lake query, $C = \{C_1, C_2, \ldots, C_n\}$ $Q$ clauses, $X = \{X_1, X_2, \ldots, X_n\}$ coverage sets such that $X_i \subseteq F$ is a coverage set of a data lake query whose "where" condition contains clause $C_i$ only.

$$\forall X' \subseteq X, Cov(\bigcap_{x \in X'}, Q) \qquad (4)$$

***Proof of Theorem 1:***

Let $Q$, $C$, $X$ be as defined in Theorem 1, and let $X' \subseteq X$. We want to prove that $S = \bigcap x \in X'$ is a coverage set of $Q$.

According to the defiition of coverage set (Definition 2), we need to prove that:

1) $S \subseteq F$
2) $\forall f \in F \setminus S, \neg \exists t \in f, S(P_Q, t)$

To prove (1), let us recall that $S$ is intersection of sets that are subsets of $F$, and hence $S$ must be a subset of $F$ as well.

To prove (2), let us assume for the sake of contradiction, that exists $f \in F \setminus S$ and $t \in f$ such that $S(P_Q, t)$. That means that $t$ is part of the $Q$ result, and hence, it satisfies each of the coverage sets in $X$ and in particular of those in $X'$, and that is a contradiction to $f$ being chosen from $F \setminus S = F \setminus \bigcap x \in X'$.

∎

***Example 1:*** As an example, let us consider the following query (Query 4) and let us assume that its estimation values are given in Table 3 (we assume here a single plan per clause and that the results are estimated based on the hypothetical extension of Table 1). Now, we are interested in finding a subset of $\{C_1, C_2, C_3\}$ that will provide us with the optimal total query execution performance. All the possible combinations are listed in Table 4 along with their estimated total cost. We can see that the optimal cost(=4) is achieved by clauses $\{C_2, C_3\}$ with the coverage set {file051}.

```
SELECT  *
FROM    metrics_table
WHERE   ( date = '2020-02-20' OR date = '2020-03-13' )
        AND val < 7
        AND metric = 'cpu'
```

*Query 4*

Table 3 Coverage Plans Estimations for Query 4

| clause | estimated cost | estimated result |
|---|---|---|
| $C_1$ = "date = '2020-02-20' OR date = '2020-03-13' " | 5 | {file170, file051} |
| $C_2$ = " val < 7" | 1 | {file051, file033, file302, file048} |
| $C_3$ = "metric = 'cpu'" | 2 | {file201, file170, file051, file079} |

Table 4 Coverage Plans Combinations for Query 4

| clauses | estimated result | estimated total cost |
|---|---|---|
| $\{C_1\}$ | {file170, file051} | 2 + 5 = 7 |
| $\{C_2\}$ | {file051, file033, file302, file048} | 4 +1 = 5 |
| $\{C_3\}$ | {file201, file170, file051, file079} | 4 + 2 = 6 |
| $\{C_1, C_2\}$ | {file051} | 1 + 5 + 1 = 7 |
| $\{C_1, C_3\}$ | {file170, file051} | 2 + 5 + 2 = 9 |
| **$\{C_2, C_3\}$** | **{file051}** | **1 + 1 + 2 = 4** |
| $\{C_1, C_2, C_3\}$ | {file051} | 1 + 5 + 1 + 2 = 9 |

To summarize this example, based on the given estimations in Table 3, the optimal execution strategy for Query 4 is:
1. Get coverage set of $C_2$ (val<7) with the estimated 1 cloud read.
2. Get coverage set of $C_3$ (metric='cpu') with the estimated 2 cloud reads.
3. Perform the query only on the files in the intersection of results from (1) and (2). The expected number of files is 1. So, in total, we perform around four reads from the cloud instead of |F|.

So, intuitively, we are looking for a balanced solution, where we balance between the estimated coverage plan cost and its result size. The following "Balanced Query Coverage Plan Problem" defines this problem formally (see Fig. 4 to review the terminology).

**Definition 6** *(Balanced Query Coverage Plan Problem (BQCPP)):* Given $C = \{C_1, C_2, \ldots, C_n\}$ representing query clauses, with $C_i = \{P_{i1}, P_{i2}, \ldots, P_{ik}\}$ being the set of possible coverage plans of the clause $C_i$, and $P_{ij} = (c, R)$, where $c \in N$ is the estimated cost of $P_{ij}$, and $R \in G$ is its estimated result. $(G, \otimes)$ is a semigroup where $G$ represents a domain of possible plan results and binary operation $\otimes$ defines results' intersection. A function $size : G \to N$ assigns a positive integer value to each result in $G$. The problem is to find the smallest set of plans $X \subseteq P = \cup C_i$ such that at most one plan is taken for each clause and the sum of $X$ plans' estimated costs and the *size()* of their estimated results' intersection (by $\otimes$) is minimized (we call such $X$ a balanced coverage plan):

$$X = \underset{X}{\mathrm{argmin}} \left\{ |X| : X \in \underset{Y \subseteq P}{\mathrm{argmin}} \left( \sum_{y \in Y} y.c + size\left( \bigotimes_{y \in Y} y.R \right) \right), \neg \exists i, j, k : P_{ij} \in X, P_{ik} \in X \right\} \quad (5)$$

**Definition 7** *(Coverage Plan Estimated Cost):* The estimated cost of the given coverage plan $P$ is defined as:

$$C_e(P) = \sum_{p \in P} p.c + size\left( \bigotimes_{p \in P} p.R \right) \quad (6)$$

Looking again at Example 1 above, we can define BQCPP parameters for this example as follows:
- $C = \{C_1, C_2, C_3\}$
- $C_1 = \{(5, \{file170, file051\})\}, C_2 = \{(1, \{file051, file033, file302, file048\})\}, C_3 = \{(2, \{file201, file170, file051, file079\})\}$
- $G = 2^F, \otimes$ is set intersection, *size()* is set cardinality

To achieve more accurate results, instead of estimating coverage results as files, we can estimate them at the records level. Then, $G$ would be defined as $2^T$, and *size* would be defined as $size(R) = |\{f \mid \exists t \in R, t \in f, f \in F\}|$. Similarly, we can use BQCPP in many other different variations.

BQCPP clearly gives us what we need - the optimal plan to calculate a coverage set for the given query. However, it introduces two significant challenges: first, it is not clear how we can efficiently estimate coverage files or records for a given clause; second, according to Theorem 2, it is NP-hard. We deal with both these challenges in the following sections: estimation is based on the traditional *selection size estimation* [56] adjusted to the cloud data lake architecture (Section 4.2.3.2), and NP-hardness is tackled by using heuristic approaches (Section 4.2.1).

**Theorem 2:** BQCPP is NP-hard when $\otimes$ is defined as set intersection, and *size()* is defined as set cardinality.

***Proof of Theorem 2:***

We prove by reduction from the *set covering* problem (SCP) [108] where inputs are $U = \{1, 2, \ldots, n\}$ and $S = \{S_1, S_2, \ldots S_m\}, S_i \subseteq U, \bigcup S_i = U$. The goal of SCP is to find the smallest $X \subseteq S$ such that the union of $X$ values equals $U$:

$$X = \underset{X \subseteq S}{\mathrm{argmin}} \left\{ |X| : \bigcup X = U \right\} \quad (7)$$

Let $U$ and $S$ be the inputs to SCP, we build inputs to BQCPP as follows (note that BQCPP is hard even when using a single plan per clause and the same cost for all the plans):

- $G = 2^U, \otimes$ is set intersection and *size()* is set cardinality
- $C = \{C_i | C_i = \{P_{\{i_1\}}\}, P_{i_1}.c = 0, P_{i_1}.R = U \setminus S_i, \forall i \in \{1,2,...,m\}\}$

We build the output to SCP ($X$) based on the output of BQCPP ($X'$) as follows:

$$X = \{S_i | C_i \in X'\} \tag{8}$$

Clearly, the construction of the input and the output can be done in polynomial time. Now, we need to prove that $X$ is indeed the correct output to SCP with arguments $U$ and $S$. Let us assume, for the sake of contradiction, that $X$ is not a correct output. Then, either the union of $X$ values does not equal $U$, or there exists another subset of $S$, such that the union of its values equals $U$ and its cardinality is lower than that of $X$. Below we show that both options lead to a contradiction.

1) Let us assume first that the union of $X$ values does not equal to $U$. Then, there is $u \in U$ such that $u \notin \cup X$. If $X = \{S_i, S_j, ...\}$, then $X' = \{C_i, C_j, ...\} = \{\{(0, U \setminus S_i)\}, \{(0, U \setminus S_j)\}, ...\}$. Since we know that the union of all $S$ values equals $U$, we know that the intersection of all $P_{i_1}.R$ in $\{C_i, C_j, ...\}$ from $X'$ is $\emptyset$ (BQCPP returns intersection set with the lowest cardinality). Then, $U \setminus S_i \cap U \setminus S_j ... = \emptyset$, but also there is $u \in U$ that does not belong to any of $S_i \in X$ which is a contradiction.
2) Let us assume now that exists $T \subseteq S$ such that the union of $T$ values equals $U$ and $|T| < |X|$. Let $T = \{S_i, S_j, ...\}$, then $\cup T = U \rightarrow \{U \setminus S_i \cap U \setminus S_j, ...\} = \emptyset$. Then, $T' = \{C_i, C_j, ...\}$ is a valid output of BQCPP, but $|T| < |X| = |X'|$ which is a contradiction to $X'$ being a solution to BQCPP with the minimal cardinality.

∎

A simple example that demonstrates the reduction is as follows:
- Let $U = \{1,2,3,4,5\}$ and $S = \{\{1,2\}, \{1,3,4\}, \{2,4,5\}, \{5\}\}$ be the inputs to SCP
- Input to BQCPP, $C = (0, \{3,4,5\}), (0,\{2,5\}), (0,\{1,3\}), (0,\{1,2,3,4\})$
- The response of BQCPP, $X' = \{(0,\{2,5\}), (0,\{1,3\})\}$ (the smallest subset with the minimal cost and intersection size)
- The response of SCP, $X = \{\{1,3,4\}, \{2,4,5\}\}$ (the smallest subset whose union equals to $U$)

The reduction is based on the following observation: If we have a subset of S such that the union of its values equals U, then the intersection of the complements of its values is equal to ∅ (and vise versa).

Now we can outline the high-level architecture of our overall solution (Fig. 5):
1. Client submits a query to the query engine.
2. Query engine passes the query to our "Optimization Framework" (OF), which is denoted by the green dotted square in the diagram. OF can be embedded in the query engine or run as a separate service.
3. OF consists of the following three modules (marked by red numbers in the diagram):
    a. *Estimations calculation* - this module receives the original query from the query engine and prepares estimation values in the format of BQCPP input defined in Definition 6 (and depicted in Fig. 4).
    b. *BQCPP Solver* - this module receives BQCPP input from module 1 and solves the BQCPP problem to find a balanced coverage plan $P$. We want to be sure that OF improves query performance and does not degrade it, so we check that the balanced coverage plan $P$ has an estimated cost below the predefined threshold $K$ (if it does not, we fallback to the default flow).
    c. *Coverage plans execution* - module 3 gets the balanced coverage plan $P$ from module 2 and executes it to get the corresponding coverage set (each one of the plans from $P$ is executed in parallel, and their results are intersected) and returns it to the query engine.
4. Query engine reads the coverage set files from the storage, executes the query based on the retrieved files, and returns the query result to the client.

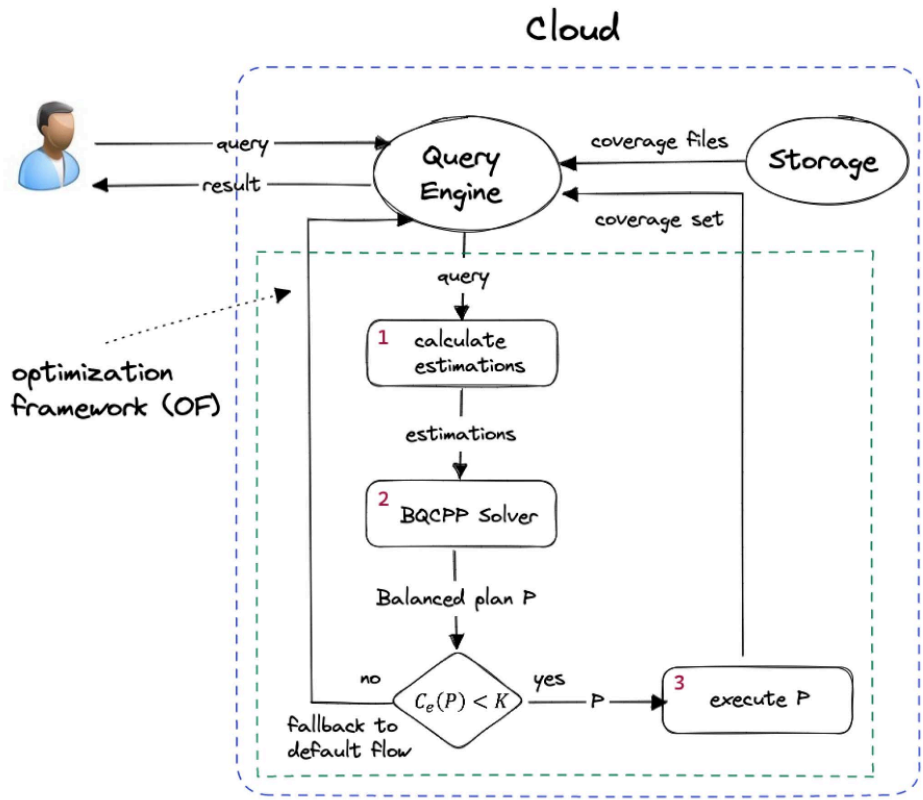

*Figure 5 High Level Architecture of the Indexing Scheme*

Our architecture is based on the following two core principles:
- **Do no harm:** We want to be sure that we do not degrade existing query performance; we can either improve it or keep it as-is. In our optimization framework, modules 1 and 2 should have all the required information in-memory and hence, in our cost model, can not increase the cost of the query. The only component that can increase the cost is module 3, but it is cost-bound, meaning that users can limit the maximal estimated cost of the balanced coverage plan, and in case this threshold ($K$) is reached, fallback to the default (existing) strategy. An obvious threshold is $|F|$, but because the cost is estimated, different users can use different thresholds according to their datasets and needs.
- **Pluggability:** It can be seen that our architecture is generic and agnostic to the actual types of the different components. We can run on any cloud, use any query engine, storage layer, etc. New components introduced in our optimization framework (BQCPP algorithms, coverage plans, estimations) also can be replaced by more efficient (or appropriate) implementations.

In the following sections, we explain how we implement each one of the OF modules. First, we explain how we solve BQCPP (Section 4.2.1); then how we calculate coverage sets (Section 4.2.2); and finally, how we calculate estimation values (Section 4.2.3). It is important to note here that our implementation of the modules is only one of the many possible options. Our architecture allows replacing each of the modules with another implementation and even mixing different strategies in the same module. The only constraint is to respect the following basic contract:
- Module 1 gets a query and returns corresponding estimations as BQCPP input (Definition 6).
- Module 2 gets BQCPP input and returns BQCPP output (Definition 6).
- Module 3 runs the given coverage plan and returns the corresponding coverage set.

## 4.2 Optimization Framework

### 4.2.1 BQCPP Solver (Module 2)

BQCPP is NP-hard when $\otimes$ is defined as set intersection, and size() is defined as set cardinality (Theorem 2). To deal with its hardness, we propose the following two simple strategies:
1. *Optimistic (Algorithm 2):* We assume here that in most real-world scenarios, the number of clauses and plans is relatively small (at most dozen clauses[2] per query with one or two plans[3] per clause). In such scenarios, we can simply iterate through all the options (meaning all the subsets) and choose the best one (just like we did in the example presented in Table 4 above). Since the number of options is fixed, the running time is *O(1)*.

---
[2] Industry benchmarks (e.g. [57]) that simulate real-world workloads rarely exceed 10-30 filtering conditions in a single predicate.
[3] Plans here correspond to the different algorithms for calculating coverage sets (Section 4.2.2) so we do not expect more than a handful number of them.

2. *Greedy (Algorithm 3):* For predicates with many clauses and/or plans, we can perform a simple greedy algorithm: Iterate over all the plans (lines 4–12), while considering at most one plan per clause, and on each iteration add a plan to the result set if it produces the minimal total cost (line 14). $T_1$ denotes the current plan, and $T_2$ represents the potential new plan after adding one additional clause. We stop when no plan can improve the cost (line 16) of the current set or the plans for all the clauses were added (line 18). In the worst case we iterate through all the clauses and for each clause through all the plans, so the complexity of the greedy algorithm is $O(m^2 k)$ (where *m* is the number of clauses and *k* is the number of plans per clause). In the first iteration, we scan *m* clauses; in the second, we scan *m-1* clauses, and so on. Considering the example estimations in Table 3 again, the greedy algorithm would start with the clause that has the minimal total cost ($C_2$), then it would add $C_3$ as it reduces the cost from 5 to 4, and then it would stop as adding $C_1$ would increase the cost from 4 to 9.

---

**Algorithm 2:** Optimistic BQCPP Solver.

**Input:** clauses $C$ with the associated plans and estimations, functions $\otimes, size()$
**Output:** balanced coverage plan

1: $out \leftarrow \emptyset$
2: **for** each $X \subseteq \bigcup C$ s.t. $\neg \exists i, k, j : P_{ij} \in X, P_{ik} \in X$ **do**
3:     $\text{cost} \leftarrow \sum_{x \in X} x.c$
4:     $\text{size} \leftarrow size(\otimes_{x \in X} x.R)$
5:     $\text{sum} \leftarrow \text{cost} + \text{size}$
6:     **if** $out = \emptyset$ or $\text{sum} < \sum_{x \in out} x.c + size(\otimes_{x \in out} x.R)$ **then**
7:       $out \leftarrow X$
8:     **end if**
9: **end for**
10: return $out$

*Algorithm 2*

**Algorithm 3:** Greedy BQCPP Solver.

**Input:** Clauses $C$ with the associated plans and estimations, functions $\otimes, size()$
**Output:** approximated balanced coverage plan

```
1:  out ← ∅
2:  repeat
3:      T₁ ← out
4:      for each P_{ij} ∈ ⋃C s.t. ¬∃P_{ik} ∈ out do
5:          T₂ ← out ∪ P_{ij}
6:          cost ← ∑_{x∈T₂} x.c
7:          size ← size(⊗_{x∈T₂} x.R)
8:          sum ← cost + size
9:          if T₁ = ∅ or sum < ∑_{x∈T₁} x.c + size(⊗_{x∈T₁} x.R)
            then
10:             T₁ ← T₂
11:         end if
12:     end for
13:     if ∑_{x∈T₁} x.c + size(⊗_{x∈T₁} x.R) < ∑_{x∈out} x.c +
        size(⊗_{x∈out} x.R) then
14:         out ← T₁
15:     else
16:         break
17:     end if
18: until |out| < |C|
19: return out
```

*Algorithm 3*

### 4.2.2 Coverage Sets Calculation (Module 3)

Here, we focus on module 3 from Fig. 5. We assume that we got from the previous module (2) a balanced coverage plan $P$, and we need to execute it to get the corresponding coverage set. Since we present a single strategy, we do not need the granularity of plans and can focus on clauses only. Let $C = \{C_i, C_j, ...\}$ be clauses associated with the given coverage plan $P$. Our goal is to find the corresponding coverage set. Our approach is based on indexing. We first explain how we build and update our indexes (Section 4.2.2.1) and then how we use them to calculate query coverage sets (Section 4.2.2.2).

4.2.2.1 Index Computation

We start with the following definition of the data lake index.

**Definition 8** *(data lake index):*

$$I_{c_i} = \{ (v, j, k) | \exists t_k \in f_j, f_j \in F, (c_i, v) \in t_k, c_i \in L, v \in D_i\} \qquad (9)$$

Less formally, we can say that the *data lake index* on column $c$ in table $T$ is a relation with the following three columns:

- value - all non-empty values of $c$ in $T$ ($v$ in Definition 8)
- file - file id where a particular column value appears ($j$ in Definition 8)
- record [4] - record id where a particular column value appears ($k$ in Definition 8).

Using this definition, indexes built on columns "value" and "date" from Table 1 would look as in Table 5. Our index maps column values to their records. This design allows us to calculate the tight coverage set for any given query; however, it requires relatively high storage overhead. For some scenarios (e.g., needle in a haystack [1]), it may be sufficient to use the index only at the file level, which would require much less storage.

*Table 5 Index Example*

| | $I_{val}$ | | | $I_{date}$ | |
|---|---|---|---|---|---|
| value | file | record | value | file | record |
| 6 | 051 | 7 | 2020-02-10 | 201 | 1 |
| 8 | 170 | 4 | 2020-02-14 | 201 | 2 |
| 11 | 201 | 3 | 2020-02-16 | 170 | 4 |
| 47 | 201 | 1 | 2020-02-18 | 201 | 3 |
| 58 | 201 | 2 | 2020-02-20 | 170 | 5 |
| 66 | 170 | 6 | 2020-02-21 | 170 | 6 |
| 71 | 051 | 9 | 2020-03-13 | 051 | 7 |
| 88 | 170 | 5 | 2020-03-22 | 051 | 8 |
| 92 | 051 | 8 | 2020-03-28 | 051 | 9 |

---

[4] For the sake of the formulation simplicity, we use global record id (in a table level) to define the "record-id" in Definition 8, but record id in a file-level would be sufficient as well.

We store index tables in the data lake as regular tables sorted by the indexed column value and partitioned by column name. To speed up index queries, we build the root index - a table that serves as an index of indexes. The root index stores per each index file a single record that summarizes the most important statistics about the file. The creation of the root index can be represented by Query 5 below. An example of the root index for the indexes in Table 5 might look as in Table 6. The root index allows us to efficiently locate relevant index files by checking their min/max values. We assume that the root index is cached by the query engine. All index files are stored in a standard columnar format.

Creation of index files, as well as the root index update, is performed by the same ETL that inserts new bulk of data into the lake (step 1 in Fig. 1). Just like in relational databases, we create indexes only for searchable columns defined by a user and only those that have a relatively high cardinality.

```
CREATE TABLE root_index AS
  (SELECT indexed_column_name,
          index_filename,
          Max(value),
          Min(value),
          Count(*),
          Count(DISTINCT value)
   FROM   index
   GROUP  BY indexed_column_name,
             index_filename)
```

*Query 5*

*Table 6 Root Index Example*

| col  | file    | min        | max        | cnt | cntd |
|------|---------|------------|------------|-----|------|
| val  | index11 | 6          | 47         | 4   | 4    |
| val  | index12 | 58         | 92         | 5   | 5    |
| date | index21 | 2020-02-10 | 2020-02-20 | 5   | 5    |
| date | index22 | 2020-02-21 | 2020-03-28 | 4   | 4    |

4.2.2.2 Index Usage

Algorithm 4 presents a function that computes the tight coverage set of the query with the given set of clauses $C$, assuming the availability of data lake indexes ($I_{c_i}$) for each column appearing in the clauses. Note that we compute the tight coverage of the subset of the original query clauses and not the tight coverage of the original query (we rely here on Theorem 1), and hence we do not need to build indexes on all columns that appear in the "where" condition, only the most relevant ones. Algorithm 4 can be easily parallelized (our implementation based on Apache Spark [23] is available in [55]).

Algorithm 4 iterates over the clauses (line 2) and clauses' terms (line 4) for which we have data lake indexes, and for each term, computes a relation representing files and records satisfying the corresponding term (lines 6, 9). Files and records are computed based on relevant indexes. For terms of type <column **op** value>, we simply apply the given operation and value on the corresponding index; for terms of type <column1 **op** column2>, we join between two corresponding indexes on columns "value" and "record" (value is joined by applying "op" while the record is joined by equality). The results are unioned between terms in the same clause and intersected between different clauses (line 12). Note that on the first iteration, the *result* equals $U$ (universe), which is a relation containing all possible file names and their record IDs.

**Algorithm 4:** Get Tight Coverage By Index.
**Input:** clauses $C$ of Query $Q$, data lake indexes $I_{c_i}, i \in \{1, \ldots, m\}$
**Output:** $TC(Q)$
1: $result \leftarrow U$
2: **for** each clause $c$ in $C$ **do**
3:    $temp \leftarrow \emptyset$
4:    **for** each term $t$ in $c$ **do**
5:      **if** t is of type $<c_i$ op $v>$ **then**
6:        $temp \leftarrow temp \cup \pi_{file,record}(\sigma_{\text{value op } v_i}(I_{c_i}))$
7:      **end if**
8:      **if** t is of type $<c_i$ op $c_j>$
9:        $temp \leftarrow temp \cup \pi_{file,record}(I_{c_i} \bowtie_{value,record} I_{c_j})$ **then**
10:     **end if**
11:    **end for**
12:    $result \leftarrow result \cap temp$
13: **end for**
14: **return** $\pi_{file}(result)$

*Algorithm 4*

```
SELECT *
FROM   metrics_table
WHERE  ( date = '2020-02-20' OR date = '2020-03-13' ) AND val > 90
```
*Query 6*

Let us demonstrate how based on Algorithm 4 we can compute the tight coverage set of Query 6 above (based on data from Table 1):
- for the term "date=2020-02-20", $temp_1 = \{(170,5)\}$
- for the term "date=2020-03-13", $temp_2 = \{(051,7)\}$
- for the term "val>80", $temp_3 = \{(170,5), (051,8)\}$
- final result is $(temp_1 \cup temp_2) \cap temp_3 = \{(170,5)\}$
- hence the tight coverage of Query 6 is $\{170\}$

### 4.2.3 Estimations (Module 1)

As an input to BQCPP Solver (module 2), we need to provide the estimated cost and estimated result of Algorithm 4 for each of the clauses in the given query. As mentioned above, we can ignore clauses containing non-indexed columns (Theorem 1), so we set the estimated cost for all such clauses as $+\infty$ and their estimated result as $F$. For the rest of the clauses, we perform the following calculation.

#### 4.2.3.1 Cost Estimation

The cost of a particular clause $C_i = T_{i1} \vee T_{i2} \ldots \vee T_{ik}$ can be estimated precisely by the following calculation based on the root index:
1. For terms of type <col **op** val> we scan the root index to find ranges that can satisfy the term based on min/max values. The estimated cost of the term $C_e(T)$ is the number of corresponding index files.
2. For terms of type <col1 **op** col2> we need to perform a full scan of indexes $I_{col1}, I_{col2}$, so the estimated cost of the term, in this case, is the number of index files in both indexes: $C_e(T) = |I_{col1}| + |I_{col2}|$.
3. Finally, the estimated cost of disjunction of two terms $T_1 \vee T_2$ is equal to $C_e(T_1) + C_e(T_2)$. Hence, the estimated cost of the particular clause is the sum of the estimated costs of its terms - $C_e(C_i) = C_e(T_{i1}) + C_e(T_{i2}) + \cdots + C_e(T_{ik})$

#### 4.2.3.2 Result Estimation

The result estimation is based on the regular relational databases selection size estimation [56], in which we can estimate the number of records satisfying the predicate based on the statistical information (i.e., $G$ value in Definition 6 equals $N$). In our case, statistics are stored in the root index (which can serve as a histogram for the estimation needs). For example, the term of type "age = 30" can be estimated by Query 7 below.

```sql
SELECT  cnt / cntd
FROM    root_index
WHERE   col = 'age' AND 30 >= min AND 30 <= max
```
*Query 7*

Once we have the estimated number of records $R$ for the given clause, we can estimate the corresponding number of files *size(R)* as the number of non-empty bins after randomly throwing $R$ balls into $|F|$ bins:

$$size(R) = |F| \cdot \left(1 - \left(\frac{|F| - 1}{|F|}\right)^R\right) \tag{10}$$

When the result is estimated by the number of records, the intersection ($\otimes$) is calculated as:

$$\otimes (R_1, R_2) = \left\lceil \frac{R_1 \cdot R_2}{|T|} \right\rceil \tag{11}$$

where |T| is the total number of records in the table. In this case, BQCPP is not NP-hard and can be solved in polynomial time by Algorithm 3. The reason why the problem becomes simpler is that when using numbers for the results estimation and the equation above for intersection, each subset with the minimal cost must be included in the final result, and that drastically reduces the number of possible solutions.

## 4.3 Experiments

We built a prototype of our solution; the implementation is available online [55]. We used Apache Spark (3.3.0) as the main compute engine for index creation and query evaluation. We used AWS as the cloud provider, its EMR platform (6.9.0) for running Spark jobs, and S3 as a storage service. For all the experiments, we used an EMR cluster with the same hardware configuration - 10 nodes of m5.2xlarge instance type (each with 8 vCore, 32 GB memory, 128 GB EBS storage).

For the benchmark, we used the TPC-H dataset (3.0.1) [57] with a scale factor of 1TB. We generated the largest (lineitem) table with around 6 billion records and stored it in S3 in three different formats (CSV, Iceberg, Parquet) and four different settings of a number of files (10,000, 20,000, 50,000, 100,000). In total, we got 12 instances of the same table (see Table 7 for details).

*Table 7 Benchmark Tables Sizes (GB)*

| num of files | CSV | Iceberg | Parquet |
|---|---|---|---|
| 10,000 | 757 | 164 | 233 |
| 20,000 | 756 | 175 | 239 |
| 50,000 | 756 | 177 | 242 |
| 100,000 | 756 | 180 | 245 |

We then created indexes for each of the 12 tables (according to the Definition 8) on all three columns (extendedprice, shipdate, commitdate) used in our benchmark queries and stored them in S3 as 3,000 Parquet files (1,000 files per each indexed column) with the average total size of 344 GB. The root index was created as a single newline-delimited JSON file according to Query 5 with an average size of 654KB. The root index was stored in S3 and cached in the Spark cluster during the query evaluation. Index and root index creation by the dedicated Spark job [55] took around 1 hour per table.

As benchmark queries, we used TPC-H Query 1 (Fig. 6) and Query 6 (Fig. 7). We used different query arguments and additional conditions to simulate queries with various query coverage degrees. The complete details about each query used in the benchmark are available in [55].

In our evaluation, we executed queries with and without our scheme. We verified that both executions returned exactly the same result. We set the upper limit of the balanced coverage plan cost $K$ as the number of data lake files $|F|$ in all experiments. The main results of the evaluation are presented in Figs. 8, 9, 10, and 11 (full experimental details are available in [55]).

```
select
        l_returnflag,
        l_linestatus,
        sum(l_quantity) as sum_qty,
        sum(l_extendedprice) as sum_base_price,
        sum(l_extendedprice*(1-l_discount)) as sum_disc_price,
        sum(l_extendedprice*(1-l_discount)*(1+l_tax)) as sum_charge,
        avg(l_quantity) as avg_qty,
        avg(l_extendedprice) as avg_price,
        avg(l_discount) as avg_disc,
        count(*) as count_order
from
        lineitem
where
        l_shipdate <= date '1998-12-01' - interval '[DELTA]' day (3)
group by
        l_returnflag,
        l_linestatus
order by
        l_returnflag,
        l_linestatus;
```

Figure 6 TPC-H Query 1 (taken from [57])

```
select
        sum(l_extendedprice*l_discount) as revenue
from
        lineitem
where
        l_shipdate >= date '[DATE]'
        and l_shipdate < date '[DATE]' + interval '1' year
        and l_discount between [DISCOUNT] - 0.01 and [DISCOUNT] + 0.01
        and l_quantity < [QUANTITY];
```

Figure 7 TPC-H Query 6 (taken from [57])

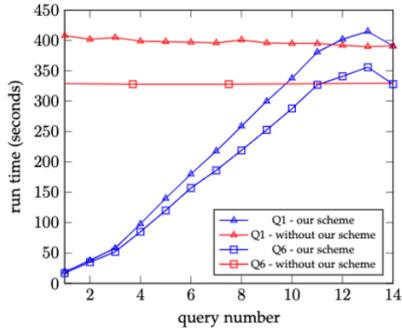
(a)
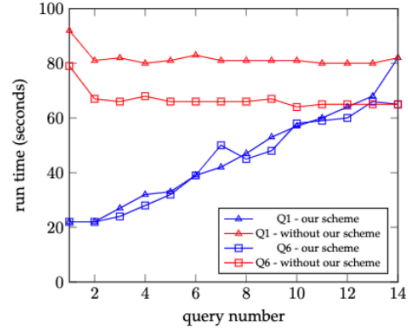
(b)
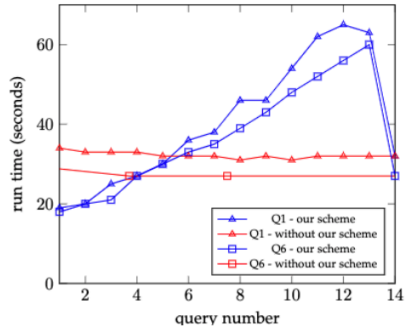
(c)

*Figure 8 Evaluation of TPC-H Q1 and Q6 on a table with 10,000 files. (a) CSV. (b) Iceberg. (c) Parquet.*

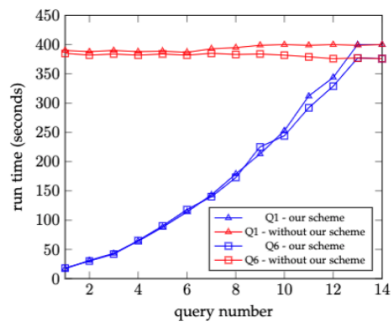
(a)
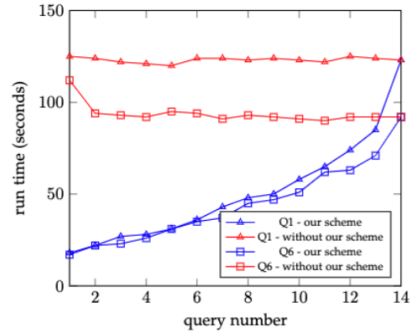
(b)
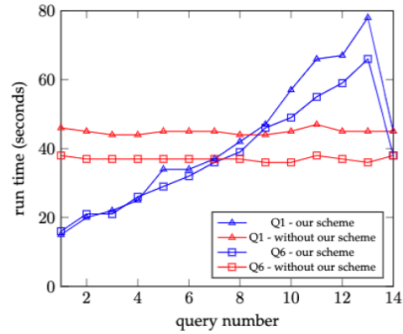
(c)

*Figure 9 Evaluation of TPC-H Q1 and Q6 on a table with 20,000 files. (a) CSV. (b) Iceberg. (c) Parquet.*

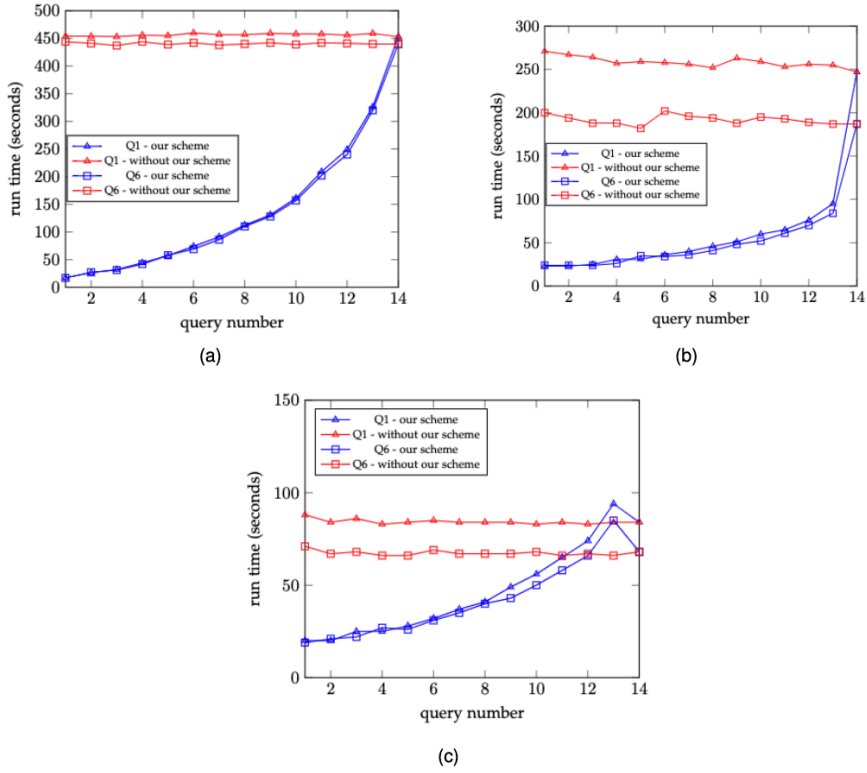

Figure 10 Evaluation of TPC-H Q1 and Q6 on a table with 50,000 files. (a) CSV. (b) Iceberg. (c) Parquet.

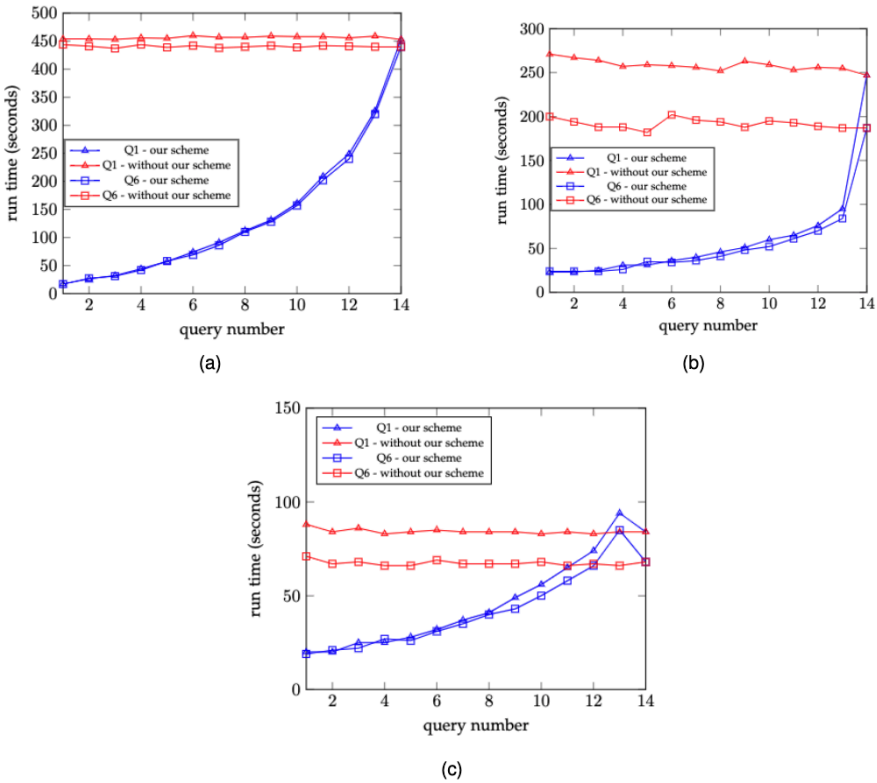

Figure 11 Evaluation of TPC-H Q1 and Q6 on a table with 100,000 files. (a) CSV. (b) Iceberg. (c) Parquet.

Graphs in Figs. 8, 9, 10, and 11 compare our scheme with the following three approaches by query run time:

- *CSV* – the baseline of the TPC-H benchmark without any optimizations discussed in Section 3.3.
- *Iceberg* – table format / lakehouse approach (Section 3.3.5)
- *Parquet* – data skipping approach (Section 3.3.2)

To better understand the trade-offs, we added the following dimensions to our evaluation:

- *TPC-H query type* – Q1 (Fig. 6) or Q6 (Fig. 7)
- *Query coverage degree* – we have 14 variations of TPC-H queries, each with a different coverage degree; the higher the query number, the higher the coverage degree
- *Number of files in the data lake* – 10,000, 20,000, 50,000, 100,000

### 4.3.1 Results Analysis and Discussion

Before starting the analysis of the experiments, let us recall the main steps of our approach (Fig. 5):

- For each query, we first extract the predicate from the "where" condition and estimate the cost and result of the coverage plan for each of the predicate clauses. We get a data structure as in Fig. 4.
- Based on the estimated values from the previous step, we find the best (balanced) coverage plan for the query. We use Algorithm 3 in the benchmark.
- If the estimated cost of the balanced coverage plan is higher than the predefined threshold $K$, we fallback to the default flow. Otherwise, we execute the plan to get a balanced coverage set for the query. Technically, we are using indexes to find the coverage set for each of the clauses in the balanced plan and then intersect their results to get the final coverage set.
- Finally, we execute the query only on the files found in the balanced coverage set from the previous step.

So, from the practical perspective, we want to achieve the following two goals:
1. Improve query execution as much as possible (by running on a subset of data lake files).
2. Do not degrade the existing query execution as much as possible (by successfully estimating the cost of our approach and falling back to the default strategy if our execution is slower).

In the below analysis, we assess to what extent we succeeded in these two goals, what are the limitations of our approach and how we can overcome them.

4.3.1.1   Analysis of query run time improvement

We have four different dimensions in our evaluation - query type (Q1/Q6), query coverage (1-14), data format (CSV, Iceberg, Parquet), and number of files (10,000/20,000/50,000/100,000). Below, we discuss the impact of each of them on the query run time.

***Query Type.*** We can see that in all graphs (Fig. 8-11), Q1 takes more time than Q6, and that is consistent across all other dimensions with and without our scheme. That result makes sense as Q6 (Fig. 7) is a simple table scan, while Q1 (Fig. 6) is a "group by" operation. Besides having different run times, Q1 and Q6 behave similarly in all other aspects, demonstrating that our approach is query-type agnostic.

***Query Coverage.*** As expected, the existing approaches (without our scheme) are almost unaffected by the query number, as they need to access all the files regardless of the coverage degree. In our scheme, on the other hand, it can be clearly seen that we perform best for queries with a low coverage degree (up to x30 times faster as in Fig. 11-a). This result is expected as our approach is based on the idea of reading only the relevant files from the data lake, and when the coverage degree is low, we need to read fewer files.

***Number of Files.*** With the dimension of "number of files", we have a similar pattern to "query coverage" but in the opposite direction. As the number of files in the data lake increases, our approach maintains more or less the same run time while other methods slow down. As an example, let us consider the performance of "Parquet" on different table sizes. The average run time increases more than fourfold when we move from a 10k table (Fig. 8-c) to a 100k table (Fig. 11-c), going from 30 seconds to 142 seconds. On the other hand, our approach only shows a slight increase in the run time, from 38 seconds to 45 seconds.

***Data Format.*** It is important to note that our approach is not intended to compete with other data formats. Rather, it is an optimization that can be applied to any of them. Our primary goal is to enhance the query performance of a particular data lake by making minimal infrastructure changes.

By analyzing the evaluation results, we can clearly see the difference between the approaches. CSV is the slowest format, while Iceberg performs much faster than CSV but slower than Parquet. Parquet is by far the fastest of the three. The slowness of CSV is expected because it is the most basic scheme, and all files are read entirely for any execution. Iceberg and Parquet store data in columnar format and can skip irrelevant columns and rows when possible. In addition, Iceberg keeps metadata that is supposed to add additional functionality and improve query performance. Surprisingly, in our benchmark, we see only the overhead introduced by Iceberg, compared to Parquet, but no improvement. The reason may be that it needs careful tuning to work best while we use the default configuration.

As a result, our scheme outperformed Iceberg and CSV in virtually all experiments. The only exceptions were queries 12 and 13 in Fig. 8-a. With regard to "Parquet," it partially outperformed us in tables with a low number of files (Fig. 8-c and 9-c). However, as the number of files grows, we perform much better. This is illustrated in Fig. 10-c and 11-c.

4.3.1.2 Analysis of query run time degradation

As demonstrated in the previous section, our approach outperforms the baseline in most experiments. Even in the small number of experiments where we run slower than the baseline, it is not a problem as long as we can anticipate it and revert to the default flow. Now, let us focus on the relevant metrics to gain insight into why, in some cases, we failed to detect the issue and what steps we can take to overcome it.

In Table 8, we have all the relevant information regarding the queries that degraded baseline performance. For each query, the table presents (by the order of the table columns):
- serial number of the entry
- data format
- number of files in the data lake ($|F|$)
- figure containing this query
- query number
- estimated balanced coverage size ($|BQCPP_E|$)
- actual balanced coverage size ($|BQCPP|$)
- number of accessed index files by the balanced coverage plan ($|index|$)

Table 8 Run time degradation metrics

| # | format | $|F|$ | Fig. | query # | $|BQCPP_E|$ | $|BQCPP|$ | $|index|$ |
|---|---|---|---|---|---|---|---|
| 1 | CSV | 10,000 | 8a | 12 | 1689 | 9515 | 25 |
| 2 | | | | 13 | 2537 | 9900 | 29 |
| 3 | Parquet | 10,000 | 8c | 6 | 386 | 4029 | 17 |
| 4 | | | | 7 | 513 | 5006 | 19 |
| 5 | | | | 8 | 578 | 6000 | 19 |
| 6 | | | | 9 | 731 | 7023 | 21 |
| 7 | | | | 10 | 781 | 8013 | 21 |
| 8 | | | | 11 | 1121 | 9032 | 23 |
| 9 | | | | 12 | 1592 | 9515 | 25 |
| 10 | | | | 13 | 2265 | 9900 | 29 |
| 11 | Parquet | 20,000 | 9c | 8 | 887 | 7425 | 18 |
| 12 | | | | 9 | 1162 | 9002 | 19 |
| 13 | | | | 10 | 1721 | 11042 | 22 |
| 14 | | | | 11 | 2140 | 13812 | 23 |
| 15 | | | | 12 | 2936 | 15364 | 25 |
| 16 | | | | 13 | 4528 | 18180 | 29 |
| 17 | Parquet | 50,000 | 10c | 13 | 13007 | 30555 | 30 |

To gain a better understanding of performance degradation, let us take a closer look at a specific query, entry #1 in Table 8. For this query our scheme estimated the result size of the balanced coverage size to be 1689 files and the number of indexes to scan in the plan was 25. Thus, the total estimated cost of the balanced coverage plan was:

$$1689 + 25 = 1714 < K = |F| = 10,000 \qquad (12)$$

As the estimated cost of the balanced coverage plan was lower than the set threshold $K$ (which we set to be equal to the value of $|F|$ in all of our experiments), we proceeded to run the coverage plan as outlined in step 3 of Fig. 5.

The actual balanced coverage plan turned out to be 9515, and the total query execution time was slower (as shown in query 13 of Fig. 8-a). Thus, we encountered two issues during this experiment. Firstly, our initial estimation was significantly off the mark (1689 vs. 9515), and secondly, even if we had predicted the actual result accurately, we would still have run our scheme and degraded performance (since $9515 < K = 10,000$).

To summarize, we have identified two problems in our approach: inaccurate estimation of a balanced coverage size and inappropriate value of threshold $K$. The first problem is not surprising since our estimation technique (Section 4.2.3.2) relies on simple statistical methods. To improve its accuracy, we can use the linear regression method. Figure 12 presents a simple approach to regressing $|BQCPP_E|$ values from Table 8 onto $|BQCPP|$, and the numeric results of applying this approach are presented in Table 9 (denoted by $|BQCPP'_E|$).

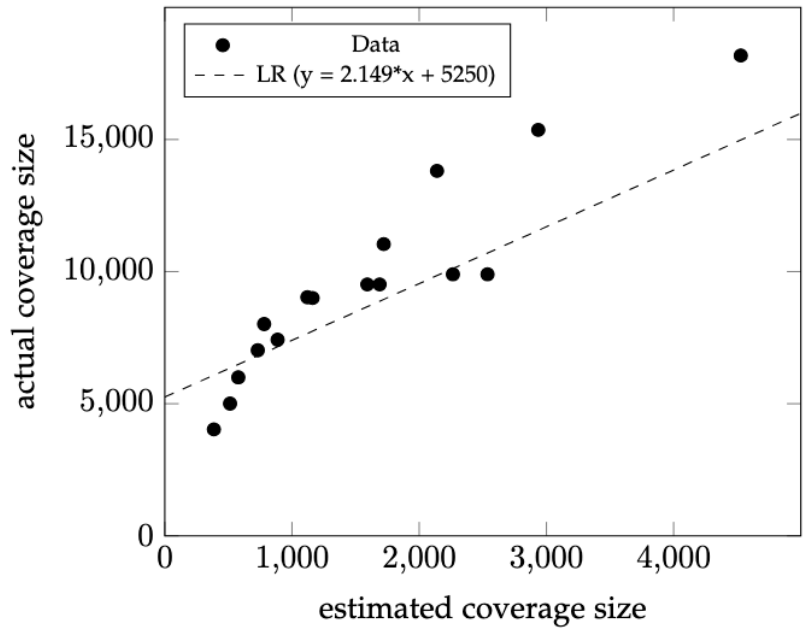

*Figure 12 Linear regression (estimated coverage size vs. actual)*

Table 9 Adjusted estimation and K threshold

| # | $K'$ | $|BQCPP_E|$ |
|---|---|---|
| 1 | 0.85 \|F\| = 8,500 | 8879 |
| 2 | | 10702 |
| 3 | | 6079 |
| 4 | | 6352 |
| 5 | | 6492 |
| 6 | 0.6 \|F\| = 6,000 | 6820 |
| 7 | | 6928 |
| 8 | | 7659 |
| 9 | | 8671 |
| 10 | | 10117 |
| 11 | | 7156 |
| 12 | | 7747 |
| 13 | | 8948 |
| 14 | 0.35 \|F\| = 7,000 | 9848 |
| 15 | | 11559 |
| 16 | | 14980 |
| 17 | 0.5 \|F\| = 25,000 | 33202 |

The second problem is also expected as setting threshold *K* to |*F*| assumes that our approach does not incur any additional overhead besides reading files from the cloud storage. However, that is not the case; one example is that we need to intersect index files after reading them, which is a costly operation. Moreover, for different data formats, the threshold should be defined differently since, for example, the read operation for CSV format is much heavier than for Parquet or Iceberg, where only a subset of the file is actually read. Similarly, the number of data lake files also affects the threshold.

Choosing the optimal value of *K* in advance is not easy, but once we have the evaluation results, we can set it properly. Table 9 shows the value of *K* (denoted as *K'*) that should be defined in each case to ensure we fallback to the default flow when needed. We have also verified that setting these values does not negatively affect other queries (those not presented in the table).

4.3.1.3   Discussion

Our experiments yield two main conclusions:
1. Our approach outperforms other schemes in most of cases, and two independent factors that make our approach faster are a low coverage degree and a high number of data lake files.
2. Inaccurate estimation of results and inappropriate threshold choice may lead to our suggestion of using a coverage plan that results in worse total query time compared to the existing approach.

Conclusion (1) demonstrates that our approach is highly relevant, especially since there are many practical scenarios where users run queries with low coverage degree, as previously mentioned in Chapter 2. In addition, real-world data lakes can contain billions of files [31], and our experiments have shown that even in data lakes with as few as 100,000 files, our scheme outperforms all other tested schemes for all queries.

Conclusion (2) is an important observation. It emphasizes the need to tune our scheme for a specific data lake to avoid query degradation. It can be done based on real user queries. While this approach is acceptable for real production systems, it would be more robust if we could eliminate this step. To achieve this, we could consider a potential solution based on the following approach: instead of assuming that we cannot access cloud storage during the estimation phase, we could "guess" that for some subsets of clauses, it is worth accessing the cloud for better estimations, which would provide us with a more accurate solution. In our future work, we plan to train machine learning models to assist us in making these "guesses".

## 4.4 Related Work

Indexing is the primary method for improving query performance in relational databases [34, 56], and it sounds reasonable to apply these well-known techniques in the cloud data lakes as well. Just like traditional indexes map column values to their actual data blocks on disk, we can map data lake column values to their files in the cloud object store, thereby finding a tight coverage set for the query.

However, whereas the general concept of using indexing in databases and data lakes is similar, some practical aspects are very different. For example:
- Big data volumes of cloud data lakes imply big data cloud indexes, and hence, highly scalable and cloud-native index implementation is required.
- Index in relational databases returns a list of record IDs that are mapped to the disk blocks, and in many cases, reading from disk via record IDs is less efficient than performing a full table scan (sequential read versus random read). The rule of thumb [34] says that "it is probably cheaper to simply scan the entire table if over 5% of the tuples are to be retrieved". In the data lake model, on the other hand, sequential reads cannot cross multiple objects, and hence, it is always better to use an index result (even if the index returned 99% of the data lake files).

These differences introduce both opportunities and challenges. On the one hand, we cannot apply existing techniques as-is due to the different scales and environments. On the other hand, we can utilize the fact that having a tight coverage set is always beneficial to design much simpler and more efficient algorithms and data structures for indexing in cloud data lakes.

Existing indexing techniques for cloud data lakes include industry approaches like [52], where a simple inverted index is stored in a key-value store. This approach, however, does not consider a relational model and focuses on very simple queries on raw data. Our previous work in [1] presents an indexing scheme for relational data in cloud data lakes where indexes are stored inside the lake, and their creation is performed by parallel algorithms. This scheme, however, limits the query model to simple selection/projection queries with a single-column predicate.

In Hyperspace [106], "covering indexes" for cloud data lakes are implemented. Covering indexes are built upon user-provided column lists "indexed columns" and "included columns," and aim to improve the performance of the queries that search by "indexed columns" and retrieve "included columns". The implementation is based on copying both "indexed" and "data" columns from the data lake and storing them in a columnar format sorted by "indexed" columns. While covering indexes can improve some types of queries, they are not suitable for queries where retrieval of all (or the majority of) the columns should be supported (e.g., for a GDPR scenario).

In Delta Lake [18], Bloom filter indexes are supported. Each data lake file can have an associated index file that contains a Bloom filter containing values of indexed columns from this file. Then, upon a user query, Bloom filters are checked before reading the files, and the file is read only if the value was found in the corresponding Bloom filter. A somewhat similar idea is implemented in [105], where an index data structure is also attached per data lake file. Here, it is not necessarily a Bloom filter and can be any appropriate index structure (B+ tree, hash index, etc'). In addition, in [105], the index is not built for all the data, but follows adaptive approach where indexes are built and destroyed based on the workload. The main difference between our approach and those presented in [18] and [105] is that their schemes build index data structures per data lake file, while our scheme builds global indexes.

Our solution can be seen as a combination of different techniques from relational databases and big data domains. Thus, we use partitions as in [27] to organize our index according to columns. For scalability, we use root-index as in B+ trees and some distributed databases (e.g. [11]). Indexes are stored in a data lake in a columnar format [39, 128], and a parallel processing paradigm [23, 24] is used for index creation and updating.

## 4.5 Conclusion

In this chapter, we address the issue of poor query performance in cloud data lakes. We use the terminology introduced in [Chapter 2](Chapter 2) to define an optimization problem (BQCPP) that finds the best (balanced) coverage set for the given query. We show that BQCPP is NP-hard in its most precise version (when the result is estimated by files or records). We deal with the hardness of BQCPP by two strategies: suggest heuristic algorithms and estimate the result by the number of records (rather than actual files or records). Our solution is based on ideas from relational databases related to indexing and statistics management. The key insight is that the storage resources are usually much cheaper than the compute resources [1, 21], so if we could provably improve query performance by adding more storage, it probably would be very useful for many data lake users. We demonstrate through the experiments that our approach works well and outperforms the existing approaches in most cases.

Our solution focuses on a very specific niche in a query optimization area - reducing the number of files read from the storage. That is why our cost model is solely based on reads from the cloud. There are many other possible optimizations, most known of which were presented in [Section 3.3](#), that are orthogonal to our work and can be easily combined with. Moreover, our scheme itself can be used for additional optimizations. For example, our index data structure can be trivially used for the optimization of index-only accesses [107], like finding min, max, count, sum, average, and distinct values. The coverage set of a join query can be computed by joining corresponding index files.

We can also consider caching parts of the index on the user's side. Previous research focused on caching within the query engine, based on the assumption that a parallel execution engine should handle each query. However, with the introduction of indexing, it is possible to execute queries end-to-end on a single user's machine for many types of queries, making cluster-based processing unnecessary. This allows caching techniques to concentrate on storing frequently accessed index sections on the user's machine, either on disk or in memory. We plan to explore all these potential optimizations in our future research.

# 5 Caching

## *Publications*

1. *Weintraub, G., Gudes, E. and Dolev, S., 2024, September. Coverage-Based Caching in Cloud Data Lakes. In Proceedings of the 17th ACM International Systems and Storage Conference (pp. 193-193).*
2. *Weintraub, G., Gudes, E. and Dolev, S., 2024, December. Predicate Containment Caching in Cloud Data Lakes. TechRxiv. (submitted to IEEE Transactions on Knowledge and Data Engineering).*

## 5.1 Overview and Intuition

Production data lakes may contain billions of files [31], and clearly, not all queries need to read all the data lake files. As shown in previous work, reading irrelevant files has a tremendous overhead on query performance [1, 4, 37, 40]. One reasonable solution to this problem is caching.

Much evidence [58-60] shows that in real-world systems, many queries have overlaps, and hence, caching some of the intermediate results and reusing them for subsequent queries makes perfect sense. Unfortunately, such an approach is problematic in a big data environment, where intermediate results might be particularly large, making traditional caching approaches impractical. In our solution, we overcome this issue by caching metadata rather than the actual data. This metadata, called *query coverage set*, was defined formally in Section 2.1 (Definitions 2 and 3). Informally, this is the set of files that needs to be read from the storage to satisfy the query.

To provide some intuition about our solution, let us look again at the example query presented in Chapter 2 (Query 1). Note that if the data lake is neither partitioned nor indexed on the predicate columns [5] ("metric" and "val"), the query engine must scan all the data lake files to find the relevant records.

Now, consider subsequent queries that need to scan a subset of the files scanned by the original query. These queries may take many forms, such as adding more AND filters to the original query, changing parameter values (e.g., increasing the "val" value), or using different selection expressions (e.g., specific columns and not "*"). Naive caching that stores the exact result of each query would not be helpful in this case since both the query and the result will be different. Storing intermediate results can help, but it is an expensive mechanism to build and maintain [59]. Therefore, in most cases, subsequent queries will need to scan all the files in the data lake again. In our approach, we aim to minimize full data lake scans in such scenarios by applying a lightweight metadata caching mechanism.

---

[5] Since partitioning can be applied only to a small subset of columns and indexing is a very heavy technique, this is a common scenario in real-world systems [59, 105]

The high-level diagram of our caching scheme is presented in Fig. 13:
1. A user submits a query $Q$ to the query engine.
2. The query engine has a cache data structure containing the previous queries' identifiers and associated coverage sets. The query engine checks if $Q$ is *contained* (defined formally below) by any of the cached queries.
3. If at least one query containing $Q$ is found in the cache, the corresponding minimal coverage set is returned.
4. If no such query was found, the query engine performs its usual flow (in the worst case, it means reading all the data lake files).
5. The query engine reads files from the storage (either by default flow or based on the coverage files found in step 3).
6. The query engine executes the query on the files read in step 5 and (optionally) updates the cache data structure with the information about the current query.
7. Finally, the query result is returned to the user.

While our scheme is straightforward, it has several research challenges:
- How do we decide if one query contains another?
- What is the best data structure for the cache in our scenario, and what are the best algorithms to query and update it?
- How do we decide what queries to store in the cache and what to evict?
- Can our solution be easily implemented in modern query engines, and can we empirically validate its efficiency and feasibility?

In the following sections, we deal with all these challenges. One of our main contributions is a new technique to efficiently answer query containment queries by mapping them to geometric space, enhancing the standard usage of spatial indexes. We believe that this enhancement is of independent interest.

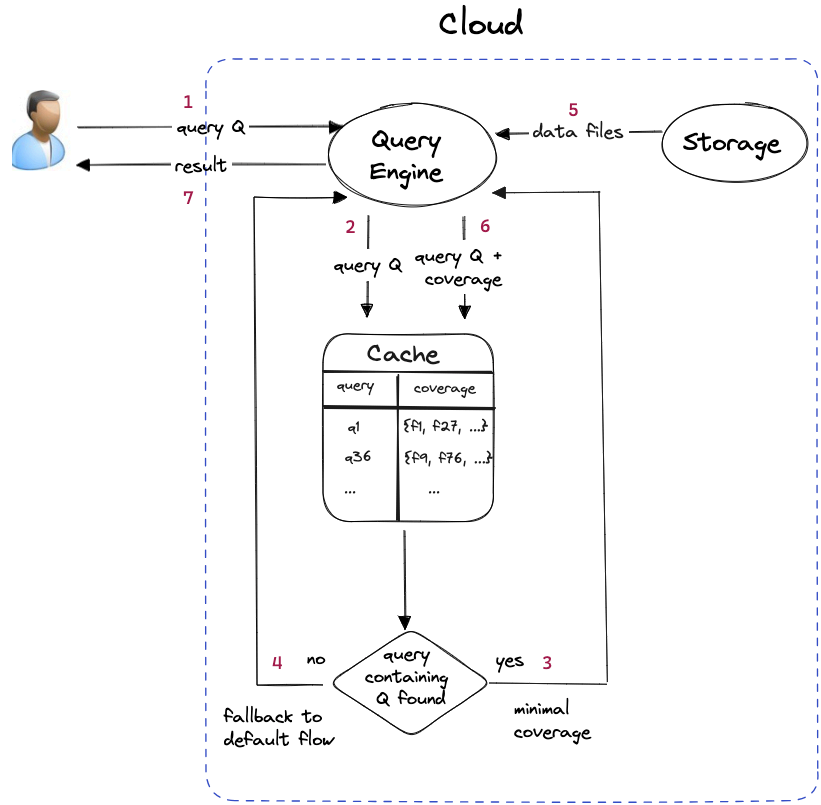

*Figure 13 Predicate Containment Caching High Level Diagram*

## 5.2 Predicate Containment Caching

### 5.2.1 Background and Preliminaries

We use the same terminology defined in Section 2.1 with some adjustments. For example, we restrict the definition of the data lake query (Definition 1) such that $P_Q$ is given as a conjunction of atomic *terms* $T_1 \wedge T_2 \wedge ...$ . A term is a condition of type <column **op** value> (e.g., age > 40). Although the choice of conjunctive predicates may appear limiting, a large number of real-world queries fall into this class [61, 62].

Note that according to this definition, each data lake query predicate can be represented by $m$ intervals (one interval for each table column). For example, the query's predicate from TPC-H Query 6 (Fig. 7) with DATE = "1994-01-01", DISCOUNT = 0.5, and QUANTITY = 25 can be represented by the $m$-dimensional interval presented in Table 10. Predicates such as "x > 5 and x > 10" are reduced to "x > 10", while predicates like "x > 5 and x < 0" are evaluated to "false".

Table 10 Predicate as an interval example

| $c_1$ | $c_2$ | $c_3$ | ... | $c_m$ |
|---|---|---|---|---|
| [1994-01-01, 1995-01-01) | [0.49, 0.51] | [$c_3.min$, 25) | | [$c_m.min, c_m.max$) |

We will denote [6] an interval representation of the predicate $P$ by:

$$I(P) = [x_1, y_1] \times [x_2, y_2] \times ... \times [x_m, y_m] \subseteq D_1 \times D_2 \times ... \times D_m \quad (13)$$

The volume of interval $I$ is defined as:

$$V(I) = \prod_{i=1}^{m}(y_i - x_i) \quad (14)$$

**Definition 9** *(query containment):* Given conjunctive data lake queries $Q_1, Q_2$,
$$Q_1 \subseteq Q_2 \leftrightarrow I(P_{Q_1}) \subseteq I(P_{Q_2}) \quad (15)$$

Our definition of query containment is different from the traditional one [130], where query Q1 is contained in query Q2 if, for any database instance, the result of Q1 is contained in the result of Q2. Since our query model is more restrictive, we can define query containment based solely on the intervals containment, allowing us to develop a new scheme that is based on this definition. For example, one important observation that stems directly from Definitions 2 and 9 is the following corollary:

**Corollary 1:** For any conjunctive data lake query Q and its coverage set X,
$$Q' \subseteq Q \rightarrow Cov(X, Q')$$

We can now define our approach (Fig. 13) based on the above semantics as follows:
- The query engine maintains a cache data structure that stores predicate intervals and tight coverages of executed queries.
- When a new query is received, the query engine checks if the given query is contained in one of the cached queries by checking if its predicate interval is contained in some of the cached predicate intervals. If found, the coverage set with the minimum number of files is returned. Based on Corollary 1, the query engine can execute the given query by reading only these files and skipping the rest.

---

[6] w.l.o.g we focus on closed intervals only; the extension to other types of intervals is straightforward

Our main goal, then, is to implement a data structure that satisfies the following requirements:
1. Supports *Put (Predicate, Coverage)* operation that stores in the cache the given query predicate and an associated tight coverage set.
2. Supports *GetMinCoverage (Predicate)* operation that returns a tight coverage set linked to the predicate containing the given predicate and has a minimal size.
3. Minimizes the cache data structure storage size.
4. Minimizes the runtime of *Put*
5. Minimizes the runtime of *GetMinCoverage*
6. Maximizes the cache hit rate.
7. Supports dynamic scenario where queries run simultaneously with files' deletion and insertion[7].

In the following section, we will introduce our caching approach, which addresses all the specified requirements. We start by presenting our basic scheme for a static scenario (Section 5.2.2.1), which satisfies only requirements (1), (2), (4), and (6). Then, in Section 5.2.2.2, we extend it to a dynamic scenario, where deletions and appends are also supported (requirement 7). In Section 5.2.3, we introduce an enhanced approach where we deal with requirements (3) and (5) as well, at the cost of slightly worsening requirements (4) and (6). Finally, in Section 5.2.4, we outline how our scheme can be extended to support queries over multiple tables.

### 5.2.2 Basic Scheme

In our basic scheme, the cache data structure is a simple linked list of pairs $(I(P), X)$, where $I(P) \subseteq D_1 \times D_2 \times ... \times D_m$ is a predicate represented by an *m*-dimensional interval, and $X \subseteq F$ is a set of file names.

We start by explaining our basic scheme in a static (read-only) mode.

#### 5.2.2.1 Static Scenario

Below, we explain how our basic read-only scheme deals with each of the requirements defined in the previous section (Section 5.2.1):
1. $Put(P, X)$ is a simple insertion into the linked list of a pair $(I(P), X)$.
2. $GetMinCoverage(P)$ – we scan the list and check if any of the intervals contain $I(P)$ and return the one with the minimal coverage size.
3. We put into the cache information about all the executed queries; hence, the storage cost is *O(n)* where *n* is the total number of executed queries.
4. The runtime of *Put* is $O(1)$.
5. The runtime of *GetMinCoverage* is $O(n)$.
6. Since information about each executed query is cached, our utilization of cache is the best possible, and the hit rate is maximized.
7. Here, we assume a read-only mode; the dynamic scenario is explained below.

---

[7] Updates are normally implemented by inserting a new, updated file and deleting the old one [18]

### 5.2.2.2 Dynamic Scenario

We need to support deletes and appends. For each delete operation, we simply scan our list and remove deleted files from all the coverage sets in the cache. For systems with a high number of deletes, we can utilize a hash table to map file names to cached entries, resulting in much faster deletes as we only need to scan the relevant cache entries rather than the entire cache.

To support appends of new files, we rely on the observation that finding a coverage set of a given query is a decomposable problem [63], meaning that we can decompose our problem into a set of smaller problems, solve them separately, and combine the results.

This idea can be explained with the following example - let us say that our data lake has two partitions $P_1 \subseteq F, P_2 \subseteq F, P_1 \cap P_2 = \emptyset$, and let us assume that our cache is built based only on $P_1$ files. Then, for any given query $Q$, we can use a coverage set computed as $GetMinCoverage(P_Q) \cup P_2$

Formally, to support a dynamic scenario, we need to make the following changes in *Put* and *GetMinCoverage:*

- *Put* - we slightly change the previous implementation (Section 5.2.2.1) by storing a triple *(I(P), X, ts),* where *I(P)* and *X* remain the same as before, and *ts* represents the maximum timestamp of the files in the data lake at the moment when *X* was computed.
- *GetMinCoverage* - Algorithm 5 presents a pseudo-code for a dynamic scenario of *GetMinCoverage*. We scan the list as before to find all the intervals containing *I(P)*, but now, for each coverage, we also add the data lake files that were added to the lake after the coverage was computed.

**Algorithm 5** GetMinCoverage

**Input:** query predicate $P$, data lake files $F$, cache data structure $Cache$
**Output:** minimal coverage set for $P$ that can be computed based on $Cache$

1: $result \leftarrow \emptyset$
2: **for** each entry $E$ in $Cache$ **do**
3:     **if** $I(P) \subseteq E.interval$ **then**
4:         $current \leftarrow E.coverage \cup F_{>E.ts}$
5:         **if** $result = \emptyset$ OR $|current| < |result|$ **then**
6:             $result \leftarrow current$
7:         **end if**
8:     **end if**
9: **end for**
10: **return** $result$

*Algorithm 5*

For systems with time-based variability, such as those following diurnal (i.e., 24-hour) patterns [125], it may be beneficial to run an entire cache update during low-demand hours. This approach can improve query performance during high-demand hours.

### 5.2.3  Enhanced Scheme

The main limitation of our basic scheme is the high storage size of the cache data structure (requirement 3) and the high run time of *GetMinCoverage* (requirement 5). Both cost *O(n)*, where *n* is the total number of queries. We are going to show now ([Section 5.2.3.1](#)) how we can improve the runtime of *GetMinCoverage* to *O(log(n))* and reduce the storage size to any predefined threshold while maximizing a cache hit ratio ([Section 5.2.3.2](#)).

#### 5.2.3.1  Improved Search

Let us recall that the main operation of *GetMinCoverage* can be formulated as: "Given a set of *m*-dimensional intervals *S* and an *m*-dimensional interval *I(P)*, find intervals in *S* containing *I(P)*". Note that this problem is related to the field of spatial databases. There is a big family of practical data structures developed for spatial databases called spatial indexes [56]. Some of the most known are R-Tree [64], KD-Tree [65], and Quad-Tree [66]. Spatial indexes are optimized for range searches of type "find all the points in a given range". The typical runtime complexity of all basic operations (insert, delete, and search) is *O(log(n))* on average. These data structures are designed for real-world applications and support high dimensions, which makes them perfect candidates for our problem. However, whether we can transform interval containment queries to range queries is not obvious. Luckily, it turns out that by using ideas from [67] and [68], we can convert our problem of *m*-dimensional interval containment into a problem of range searching in *2m* dimensions (see [Corollary 2](#) below).

**Corollary 2:** An m-dimensional interval containment query involving a set *S* of *m*-dimensional intervals and a query *m*-dimensional interval *q*, can be answered by:
1. Mapping *S* into an equal-sized set *S'* of points in *2m* dimensions, such that for each interval $[x_1, y_1] \times [x_2, y_2] \times ... \times [x_m, y_m]$ in *S* we put into *S'* a point $(x_1, x_2, ..., x_m, -y_1, -y_2, ..., -y_m)$
2. Mapping a query interval $q = [x_1, y_1] \times [x_2, y_2] \times ... \times [x_m, y_m]$ into $q' = ((-\infty, \ x_1], (-\infty, \ x_2], ..., (-\infty, \ y_1], ..., (-\infty, \ y_m])$
3. Solving the range query for *S'* and *q'*.
4. Interpreting the answer for the range query as the answer for the interval containment for *S* and *q*.

Let us demonstrate how the mapping in [Corollary 2](#) works with a simple example where 1-dimensional intervals are used (see [Fig. 14](#) below):

1) Let us say that we have a set of one-dimensional intervals $S = \{[1, 4], [3, 6], [2, 8], [6, 9], [1, 10]\}$ and a query interval $q = [5, 7]$. Our goal is to find intervals in $S$ containing $q$ (the answer is [1, 10] and [2, 8]).
2) According to step 1 in the corollary, we map one-dimensional intervals in $S$ into 2-dimensional points $S'$, such that $S' = \{(1,-4),(3,-6),(2,-8),(6,-9),(1,-10)\}$. The points are presented in Fig. 14.
3) According to step 2 in the corollary, we map a query interval $q$ into a range query $q' = ((-\infty, 5], [-\infty, -7])$. Range query $q'$ is looking for points in $S'$ whose $X$ coordinates belong to $(-\infty, 5]$ and $Y$ coordinates belong to $[-\infty, -7])$. The range of $S'$ that contains the two points of $q'$ is shown in red in Fig. 14.
4) After solving the range query for $S'$ and $q'$ (step 3 in the corollary), we get the points $(2, -8)$ and $(1, -10)$, and when we map them back to intervals (step 4) we get the expected answer to the interval problem - [2, 8], [1, 10].

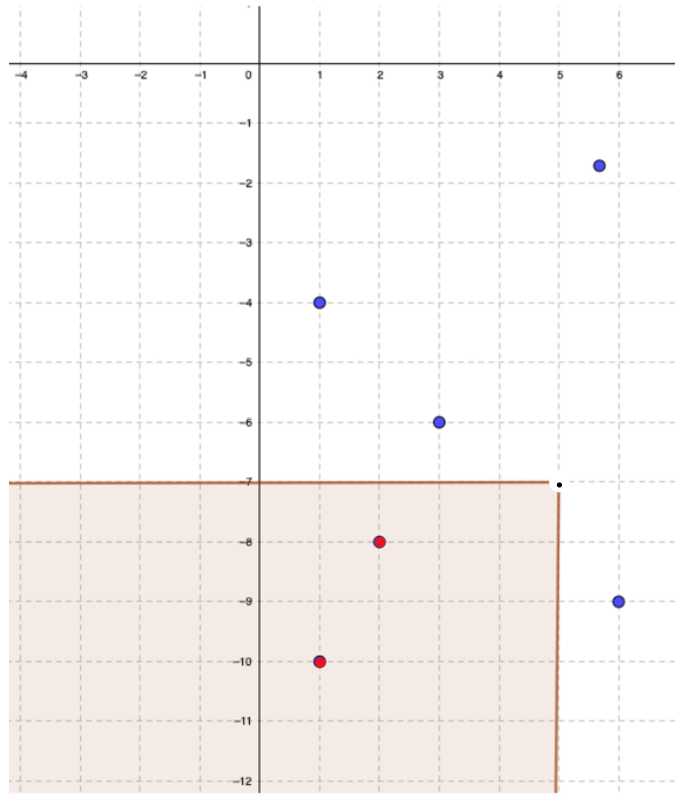

Figure 14 Corollary 2 mapping demonstration

The same idea can be easily applied to higher dimensions. For example, for 2-dimensional intervals, the mapping would look as follows:

1) $S = \{([1, 4], [3, 6]), ([2, 8], [6, 9])\}$
2) $q = ([2, 3], [4, 5])$
3) $S' = \{(1,3,-4,-6),(2,6,-8,-9)\}$
4) $q' = ((-\infty,2),(-\infty,4),(-\infty,-3),(-\infty,-5))$

5) Running a range query $q'$ on $S'$ means finding points in $S'$ where the first coordinate is in $(-\infty, 2)$, the second is in $(-\infty, 4)$, and so on.

To summarize, our improved search scheme is based on the basic scheme with the following adjustments:
- *Put* - Instead of storing *m*-dimensional intervals in a linked list, we store *2m*-dimensional points in a spatial index as explained in [Corollary 2](#). This change worsens *Put* complexity from $O(1)$ to $O(log(n))$ on average. Storage size remains the same - $O(n)$.
- *GetMinCoverage* - Instead of scanning the linked list we convert the given *m*-dimensional interval into a *2m*-dimensional interval as explained in [Corollary 2](#) and run a range query against the spatial index. Out of the retrieved points, which represent cached intervals containing the query interval, we return the minimal coverage (same logic as in [Algorithm 5](#)). The improvement in runtime is from $O(n)$ to $O(log(n))$ on average.

There are various spatial index implementations (e.g., [64, 65, 66, 69, 70]), each with its own advantages and drawbacks. In Section 5.3, we use available implementations to assess them in our scenario and identify associated trade-offs.

### 5.2.3.2 Improved Storage

So far, we assumed we could cache information about all the executed queries, and this assumption might be valid in many cases, as even millions of queries can be easily cached in memory. However, at least in some cases, we need to consider a scenario where the cache data structure is full, and we need to decide what entry should be removed from the cache to make room for the new data.

Note that from all the cached entries, we prefer those whose intervals cover the largest area (to maximize the cache hit ratio) and have the smallest coverage set size (to minimize the query cost). Based on this intuition, we can formally define what is the best way to limit our cache data structure. Given a set $S$ of $n+1$ cache entries *(P, X)* where $P$ is a predicate represented by an *m*-dimensional interval, and $X$ is a set of file names (as defined in [Section 5.2.2](#)), we are looking for $S' \subseteq S$, such that:

$$|S'| = |S| - 1 = n \tag{16}$$

$$Coverage_{total} = \sum_{i=1}^{n} |S'_i.X| \text{ is minimized} \tag{17}$$

$$V_{total} = \sum_{J \subseteq \{1,\ldots,n\}} (-1)^{|J|+1} \cdot V\left(\bigcap_{j \in J} S'_j.P\right) \text{ is maximized} \tag{18}$$

While it is easy to find *S'* satisfying Equations 16 and 17 (for example, by keeping a pointer to the entry with the maximal coverage size and removing it when needed), there is no simple solution for Equation 18. Note that because the cached intervals may overlap, we must use the inclusion-exclusion principle in Eq. 18 for the total volume calculation, and this approach is impractical due to the exponential number of terms.

Below, we suggest a few practical heuristic methods for cache replacement policy, which are evaluated in Section 5.3.

1. *Coverage-optimized* - we maintain a max-heap where the key is the cached entry's coverage set size, and a value is a pointer to the cached entry. When the cache is full, and we need to decide which entry to remove, we remove the one with the maximum coverage size.
2. *Volume-optimized* - similar to coverage-optimized, but now we maintain a min-heap where the key is a volume of the cached entry's interval (Eq. 14). When it is time to remove an entry, we remove the one with the smallest interval volume.
3. *Combined* - the idea is to combine the two previous techniques. To do that, we first normalize both metrics to the same scale ([0,1]) and direction (the higher the metric, the better). Then, we add weights to define priorities. Finally, we combine the metrics. Equation 19 below is the corresponding function where $w_1 + w_2 = 1$, and $V_{max}, V_{min}, Cov_{max}, Cov_{min}$ are max and min values of cached intervals' volumes and coverage sets' sizes respectively. We store the *g* result in a min-heap, similarly as above, and remove its minimal value when needed. Note that by setting $w_1 = 0$, we get the coverage-optimized policy, and by setting $w_2 = 0$, the volume-optimized.

$$g(P,X) = w_1 \left(\frac{V(P) - V_{min}}{V_{max} - V_{min}}\right) + w_2 \left(\frac{|X| - Cov_{min}}{Cov_{max} - Cov_{min}}\right) \tag{19}$$

To summarize, we can use one of the policies above when we have limited storage for our cache data structure. The main idea is to maintain a heap whose minimal value points to the next item candidate to remove from the cache. This approach allows us to limit the cache data structure size to any predefined threshold *k*. The incurred cost of the approach is as follows:

- *Put* - in addition to putting a new element into the cache, we put a new element into the heap and remove the relevant item from both the cache and the heap. The total overhead is *O(log(n))*, so we do not change the *Put* run time asymptotically.
- *Storage* - in addition to the cache data structure, we maintain a heap of the same size *O(k)*, meaning no asymptotic overhead in storage.

### 5.2.4 Multiple tables

As an example of queries involving multiple tables, consider TPC-H Query 3 ([Fig. 15](#)). In the "WHERE" clause, we see a combination of simple conditions of the form <column **op** value> and join conditions of type $T_i.col_x = T_j.col_y$. The core concept of predicate containment caching remains applicable. We can cache the tight coverage of the query as before, but the key difference now is that the coverage set will include files from different tables. This allows us to reuse this cache for various queries contained within the cached query.

Examples of such queries include the following:
1) Adding a join with another table (e.g., including "supplier" in the "FROM" clause and adding "l suppkey = s suppkey" to the "WHERE" clause).
2) Adding more join conditions with existing tables (e.g., adding "l shipdate = o orderdate" to the "WHERE" clause).
3) Adding more simple conditions with existing tables (e.g., adding "l quantity < X" to the "WHERE" clause).
4) Changing columns in the "SELECT", "GROUP BY", or "ORDER BY" clauses.

While the main idea of caching remains the same, it cannot be directly applied to our proposed scheme, as we cannot map predicates from multiple tables to intervals in the same manner as before. In our future work, we plan to explore this topic further and identify optimal solutions for supporting queries on multiple tables using our caching scheme.

```
select
        l_orderkey,
        sum(l_extendedprice*(1-l_discount)) as revenue,
        o_orderdate,
        o_shippriority
from
        customer,
        orders,
        lineitem
where
        c_mktsegment = '[SEGMENT]'
        and c_custkey = o_custkey
        and l_orderkey = o_orderkey
        and o_orderdate < date '[DATE]'
        and l_shipdate > date '[DATE]'
group by
        l_orderkey,
        o_orderdate,
        o_shippriority
order by
        revenue desc,
        o_orderdate;
```
*Figure 15 TPC-H Query 3 (taken from [57])*

## 5.3 Experiments

Let us recall the main idea of this chapter: We propose a caching mechanism for cloud data lakes, which is based on storing predicates and their tight coverage sets in memory and reusing them for subsequent contained queries (Fig. 13). We suggest a basic scheme (Section 5.2.2), which is based on a simple linked list, and an enhanced scheme (Section 5.2.3), which improves cache search time from *O(n)* to *O(log(n))* by using a spatial index. In addition, the enhanced scheme allows us to limit the cache size to any predefined threshold.

In our experiments, we want to evaluate two main scenarios:
1. End-to-end flow in a cloud environment that proves that our approach is relevant for real systems and works as expected.
2. Testing different dimensions with a focus on scalability - basic vs. enhanced, various spatial index implementations, cache replacement policies, number of executed queries, table columns, and data lake files.

While evaluating both scenarios in the cloud would be better, unfortunately, it is not feasible to do that for the second scenario. For example, the difference between linear and logarithmic run time can be seen only for a relatively large number of queries, but running such a large number of data lake queries in the cloud would take a lot of time and cost a lot of money. Therefore, we decided to run only scenario (1) in the cloud and do all the other tests on a local machine.

### 5.3.1 Cloud Tests

We implemented a prototype of our caching approach [71]. We used Apache Spark 3.3.1 as a query engine. The experiments were performed on an AWS EMR cluster (6.10.1) with 10 nodes of m5.2xlarge type (each with 8 vCore, 32 GB memory, 128 GB EBS storage).

For the benchmark, we used the TPC-H dataset (3.0.1) [57] with a scale factor of 1 GB. We generated the largest table (lineitem) and stored it in S3 as Parquet files. The table consisted of around 6 million records.

As a base benchmark query, we used TPC-H Query 6 (Fig. 7). We used random values in predefined ranges to simulate queries with different tight coverage sets. The complete details about the benchmark are available in [71].

We evaluated query execution in three different dimensions:
1. *Caching mode* - no caching, naive caching (store query result and reuse for the same predicate), our basic scheme caching (Section 5.2.2)
2. *Number of files in the data lake* - 10,000, 20,000, 50,000
3. *Number of executed queries in a row* - 10, 50, 100, 200

In total, we have $3 \times 4 \times 3 = 36$ instances of the experiment. Each of these 36 experiment types was executed three times, and the average time the query took (in seconds) was recorded. The results are presented in Fig. 16 and in Table 11 where each table cell contains three values (query time with no cache, query time with naive cache, query time with our cache).

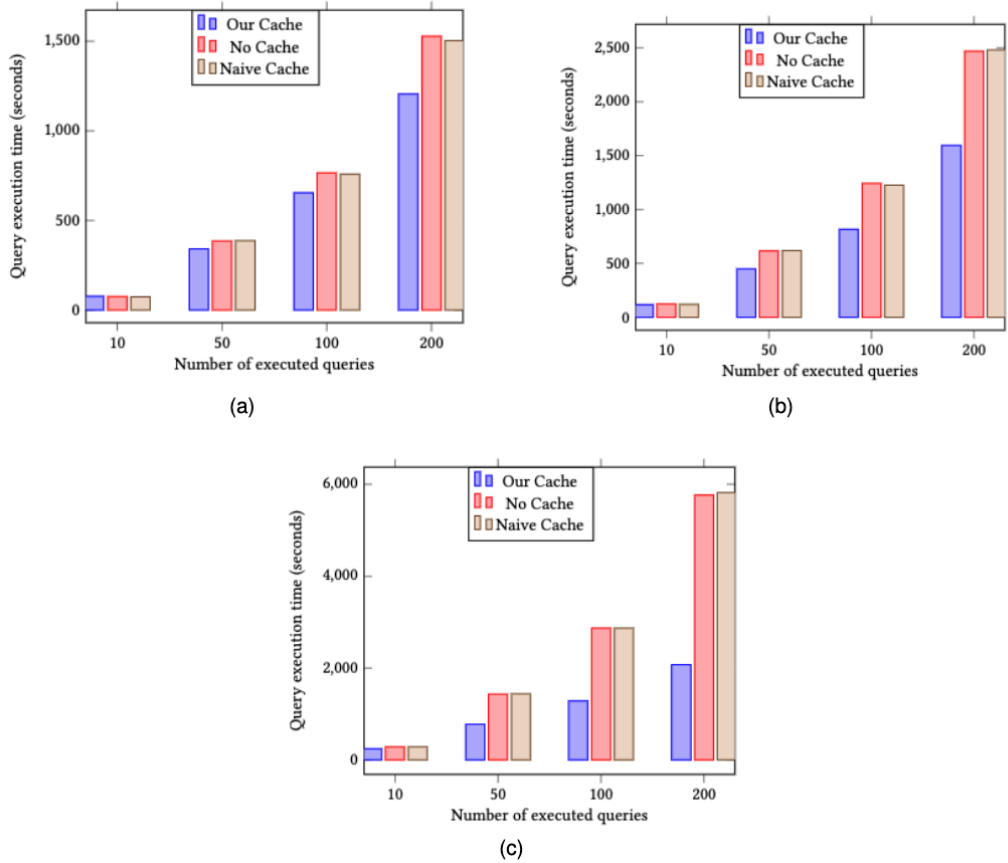

*Figure 16 Evaluation of caching scheme in the cloud with a different number of files in the data lake (a) 10k (b) 20k (c) 50k*

*Table 11 Cloud test results*

| num of files / num of queries | 10,000 | 20,000 | 50,000 |
|---|---|---|---|
| 10 | 76/75/77 | 123/121/117 | 284/284/244 |
| 50 | 385/388/342 | 617/619/449 | 1434/1439/777 |
| 100 | 765/758/655 | 1244/1226/815 | 2867/2867/1287 |
| 200 | 1528/1503/1206 | 2467/2481/1596 | 5763/5817/2076 |

5.3.1.1 Cloud Test Analysis

We can draw a few important conclusions from the cloud experiments:
- The results show that the naive cache performs almost the same as the no-cache. This is because the chance of the exact same predicate occurring is very small in a relatively small number of executed queries.

- We can see a clear query improvement in our scheme, and two independent factors that make our approach faster are a high number of queries (Fig. 17-a) and data lake files (Fig. 17-b). This result makes sense. When we execute more queries, there are more cached entries, leading to an increase in cache hits and better query performance. Similarly, when there are more data lake files, we save more time by reading specific files based on the cached coverage sets.
- As mentioned before, we could not perform too many experiments in the cloud due to time and cost constraints. However, even with our limited experiments and by using the basic scheme only, we clearly see that our approach significantly improves query execution time (up to 64% query performance improvement).

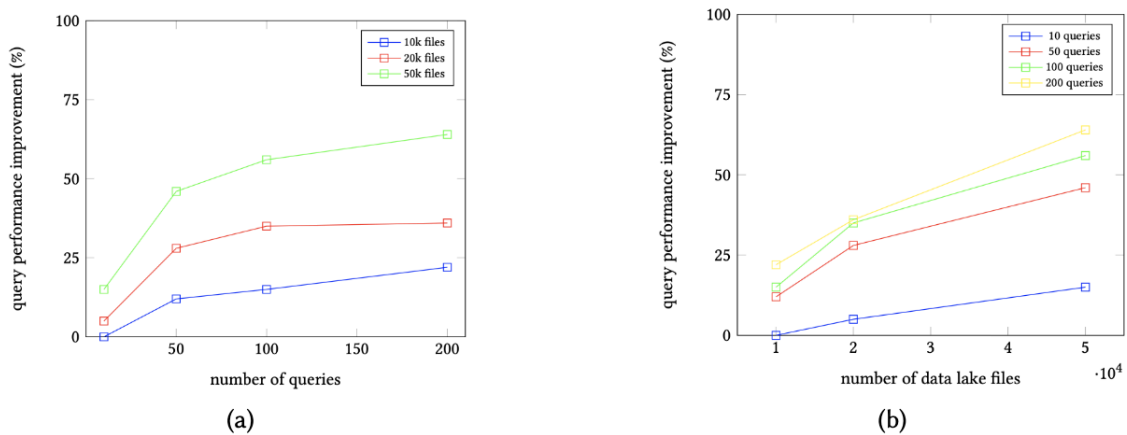

*Figure 17 Cloud query performance improvement in percentage (our cache vs. no cache). (a) by number of queries (b) by number of data lake files*

### 5.3.2 Local Tests

For our local tests, we have implemented a framework in pure Java [71] that evaluates all the components of our solution. The experiments can be easily reproduced on a stand-alone machine. We used Apple M1 Pro, with 32GB RAM, for all our tests. The framework works as follows:
1. It gets as an input: number of table columns, number of table records, number of data lake files, number of queries to execute, cache data structure (linked list or spatial index according to Section 5.2.3.1), cache replacement policy (unlimited or the one explained in Section 5.2.3.2).
2. A random table with the specified number of columns, records, and data lake files is generated. To simulate the data lake table locally, we create a hash table where keys are integers representing data lake files, and values are lists of lists with random values representing table records. Each record is randomly assigned to one of the files in a given range. When reading data from the table, we add a small delay (a few microseconds) per each table entry read that simulates the reading of files from the cloud.

3. The cache is initialized according to the defined data structure and replacement policy.
4. The specified number of random queries is performed. Queries are random in two levels - we first randomly decide how many columns will be used in a query predicate and then randomly generate predicate values. For each query execution, we record its run time and whether there was a cache hit for this query. We also perform a baseline query where no cache is used. As was shown in cloud tests, when random queries are used, the baseline results are the same for no-cache and naive cache (where query result is cached per query predicate). In fact, any caching technique that expects exactly the same query predicate (e.g., [59]) would perform similarly to our baseline.
5. Finally, we get a summary report of the test execution with the total and average query run time and cache hits.

We used our framework to perform many different tests, which can be grouped into the following three groups.

### 5.3.2.1 Table Dimensions

To evaluate different dimensions, we fixed the number of records to 1M and used different values for the number of columns, number of files, and number of queries. In this section, we used only the basic scheme.

In [Fig. 18](), we see an average query runtime per number of queries and the number of files. For this test, we fixed the number of columns at 10. This experiment confirms our findings in the cloud tests ([Section 5.3.1]()) that our scheme significantly outperforms the baseline and performs best as the number of queries and files increases.

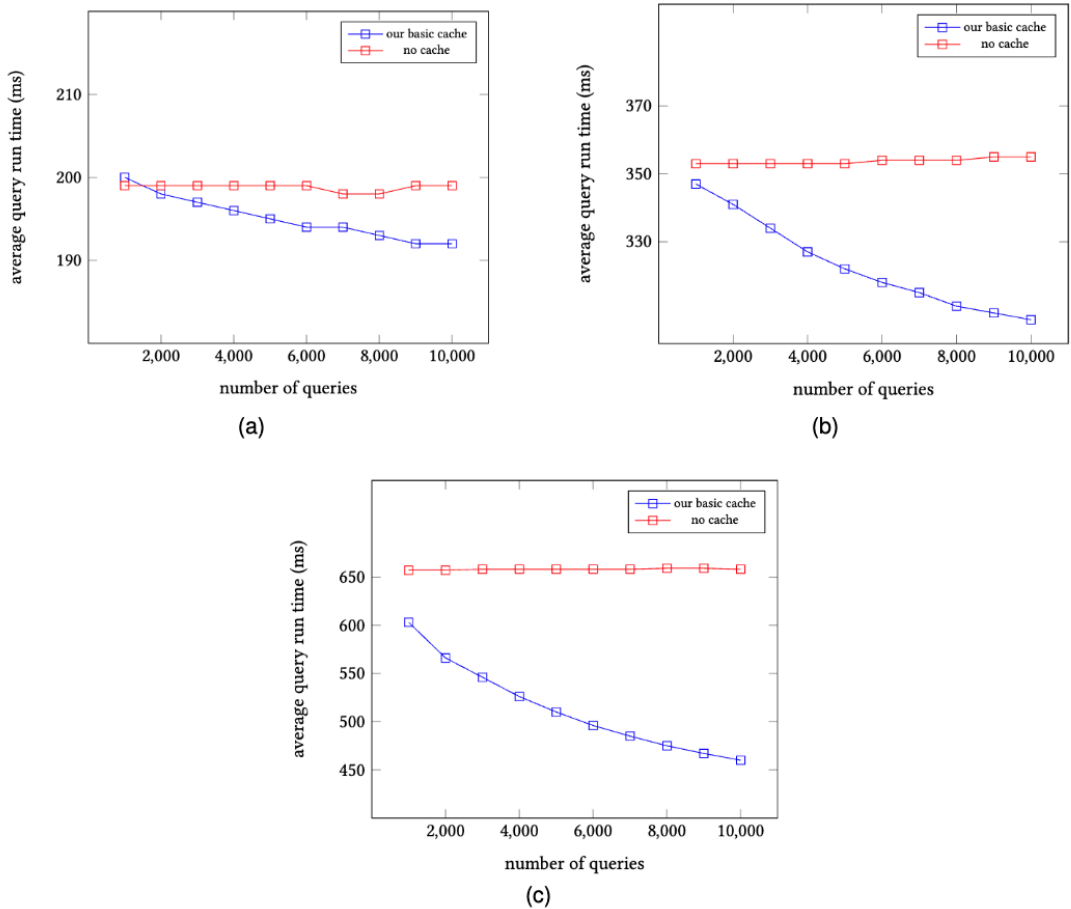

Figure 18 Local query average runtime in ms (our cache vs. no cache) by the number of queries and data lake files. (a) 50k files (b) 100k files (c) 200k files

In Fig. 19, we see an average query runtime per number of queries and the number of table columns. We fixed the number of files to 100,000 in this test. In this experiment, we can see again that as the number of queries grows, we perform better. In addition, we can see that as the number of table columns decreases, we have a better query performance improvement (up to 14% runtime improvement in Fig. 19-a vs. a mere 3% in Fig. 19-c). The reason for this result is that having many columns reduces the chance for a random query to find a query containing it, resulting in fewer cache hits.

To summarize the results so far, our experiments with the table dimensions confirm that our approach improves query performance, and three independent factors that make our approach faster are a large number of queries, a large number of data lake files, and a small number of table columns.

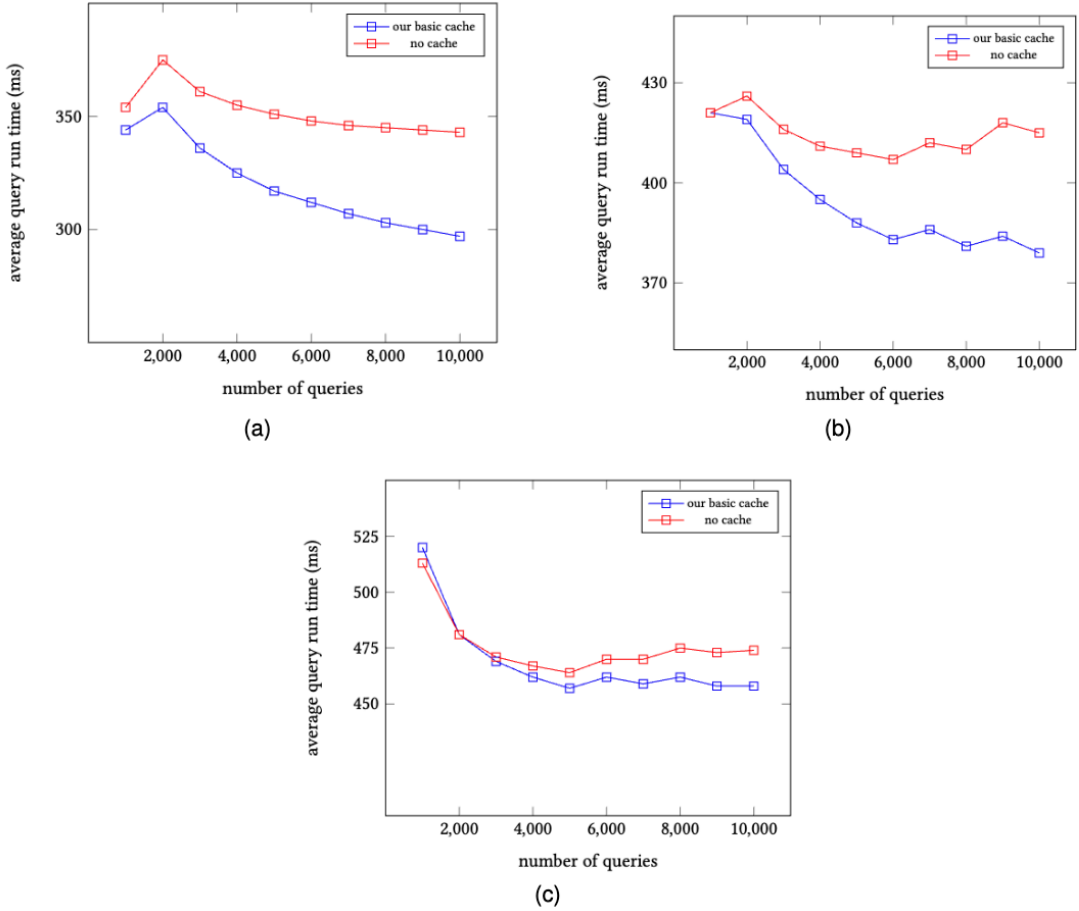

*Figure 19 Local query average runtime in ms (our cache vs. no cache) by the number of queries and table columns. (a) 5 columns (b) 20 columns (c) 50 columns*

### 5.3.2.2 Cache Replacement Policy

So far, we have not limited our cache size, and coverage sets of all executed queries were inserted into the cache. Now, we want to evaluate our cache replacement policies (Section 5.2.3.2). Our intuition was that the best results would be achieved if we balanced the coverage-optimized and volume-optimized approaches by setting different values for $w_1$ and $w_2$ constants. The reason for this expectation is that we want to cache queries with high volume (and maximize the number of queries contained in them) and low coverage size (to minimize the number of cloud reads). However, as shown in Fig. 20, the only strategy outperforming the baseline was the "volume-optimized" ($w_1 = 1, w_2 = 0$).

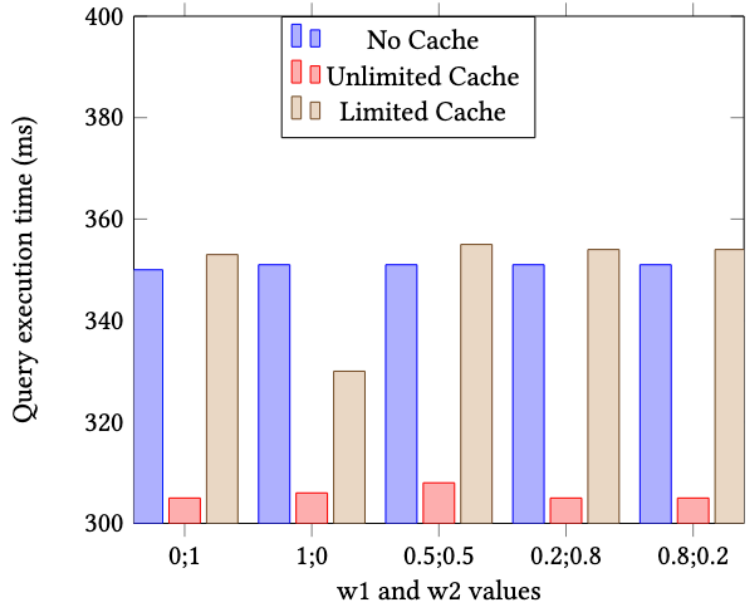

*Figure 20 Query execution time by different cache replacement policies (10 columns, 1M records, 100k files, 10k queries, cache size limited to 100 entries)*

While this result might be surprising, it makes sense. Because of the uniform distribution of the data in our experiments, we do not have cases with low coverage and high volume, and hence, when queries with low coverage are cached, they happen to be also with low volume and hence drastically reduce the cache hit ratio. It would be interesting to perform the same test in a real-world data lake where the data is (usually) not distributed uniformly.

5.3.2.3   Cache Data Structure

To evaluate our enhanced scheme, we implemented a mapping between intervals and points (Corollary 2). Then, we used available open-source implementations of various spatial indexes (R-Tree [64], Quad-Tree [66], KD-Tree [65], and PH-Tree [70]) in our experiments.

Using spatial indexes instead of a linked list for our cache data structure affects the run time of both "put" *(O(1) vs. O(log(n)))* and "get" *(O(n) vs. O(log(n)))* operations. In addition, the complexity of the linked list is given according to the worst case and spatial indexes in an average case. As the gap between the theoretical analysis and empirical performance of algorithms is a well-known phenomenon [72], our goal in this section is to confirm that experiments support our theory-driven decision.

Fig. 21 and Table 12 present the main results of this section's experiments. We used 10 columns, 100,000 records, 10,000 files, and 80,000 queries in these experiments. In the graph and the table, the total query run times per number of queries and cache data structure are shown. We can see that all tested spatial indexes outperformed both the baseline and our basic scheme based on a linked list. Out of the spatial indexes implementations, KD-Tree [65] and PH-Tree [70] gave the best results with a slight advantage of PH-Tree. We also tested Quad-Tree [66], but it constantly failed with an "out of memory" error.

*Table 12 Total query run time (seconds) by number of queries and cache data structure*

| num of queries | no cache | linked list | R-Tree | KD-Tree | PH-Tree |
|---|---|---|---|---|---|
| 10,000 | 344 | 300 | 297 | 294 | 293 |
| 20,000 | 687 | 573 | 560 | 549 | 547 |
| 30,000 | 1027 | 838 | 809 | 785 | 781 |
| 40,000 | 1370 | 1103 | 1054 | 1012 | 1006 |
| 50,000 | 1711 | 1375 | 1298 | 1234 | 1225 |
| 60,000 | 2052 | 1653 | 1542 | 1449 | 1438 |
| 70,000 | 2394 | 1936 | 1786 | 1661 | 1646 |
| 80,000 | 2737 | 2226 | 2032 | 1870 | 1854 |

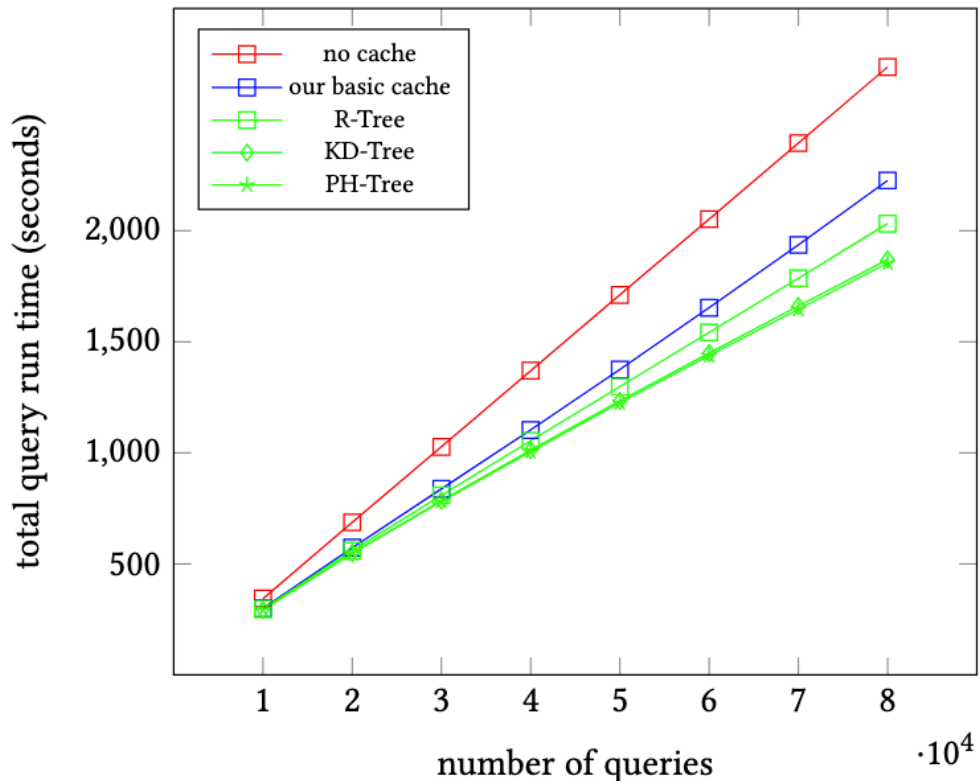

*Figure 21 Total query run time (seconds) by number of queries and cache data structure*

## 5.4 Related Work

Most query engines use some sort of caching to improve query performance. For example, in [18, 46], data retrieved from the object store is cached on compute nodes' local disks. In [28], users can explicitly cache query results in memory or on disk for subsequent usage. In Redshift [13], repeating query templates in the queries are automatically identified, and materialized views [110] to cache them are created. Then, similar queries scan the materialized view instead of recomputing the result every time [109].

An interesting concept of "separable operators" was introduced in [20]. Simply put, for some query types (e.g., projections), we can get some of the data from the cache, another part from the object store, and merge their results for the final output.

All the above caching techniques cache the actual query results, which can be problematic in a big data environment where the query results can be particularly large. A recent work of [59] suggests caching query predicates along with their "qualifying row ranges" and using them in subsequent queries with the same predicates. This technique is similar to our approach by both focusing on query predicates and caching metadata rather than data. The main difference between the approaches is that in [59], the cache is used only when exactly the same predicate is queried again, while in our approach, the match is based on predicate containment. Because of that, the approach in [59] can only improve query performance in workloads with many repeated queries, while our approach has no such limitation.

## 5.5 Conclusion

In this chapter, we presented our caching scheme for improving query performance in cloud data lakes. Our approach is based on the observation that a significant bottleneck in data lake queries is reading irrelevant files from the cloud. We suggest caching the relevant files per query (called query *tight coverage set*) and reusing them in queries contained in the cached queries. Since query containment of arbitrary queries is a hard problem [111], we limited our approach to a class of conjunctive queries that can be easily represented by multi-dimensional intervals. In future work, we want to consider a general case, for example, by using SMT solvers [112].

The most straightforward implementation of our approach is to store all executed query intervals along with their coverage sets in an in-memory linked list and try to reuse them for subsequent queries. Our tests based on the TPC-H benchmark and running on AWS cloud confirm that even such a simple approach can bring up to 64% query performance improvement.

We then show how our basic scheme can be improved in different dimensions:
- To support the dynamic scenario, we treat the problem of finding a coverage set of a given query as a decomposable problem [63].
- To improve the runtime of *GetMinCoverage*, we transform our problem

- of interval containment to a problem of finding points in a range ([Corollary 2](#)) and solve this problem by using spatial indexes.
- We analyze what is the best way to limit our cache size, and since the optimal way is computationally hard, we suggest simple heuristic techniques.

One potential extension to our work is to cache individual predicate terms separately rather than storing entire query predicates as we currently do. For example, if we have a query with the predicate "x > 20 and y < 50," we could cache "x > 20" and "y < 50" separately along with their respective tight coverage sets. This way, we could reuse the coverage of "x > 20" for another query, such as "x > 60 and z = 8."

However, a challenge with this approach is that by the end of the query processing, we may not have coverage information for each term individually. This occurs because some irrelevant results could have been filtered out by other techniques, such as partitioning, data skipping, or even our own caching methods. We plan to study this issue in our future work.

Our work has a few limitations. In addition to limiting the type of queries to conjunctive only, we also limited them to run against a single table. As discussed in [Section 5.2.4](#), in future work, we plan to extend our scheme to multiple tables and support joins. Our tests generated both the data and the queries according to a uniform distribution, and obviously, such tests can not properly simulate real-world systems. For our future work, we plan to cooperate with companies querying real-world data lakes on a scale and evaluate our approach there.

# 6 Genetic Data

***Publications***

1. *Weintraub, G., Hadar, N., Gudes, E., Dolev, S. and Birk, O., 2023, June. Analyzing large-scale genomic data with cloud data lakes. In Proceedings of the 16th ACM International Conference on Systems and Storage (pp. 142-142).*
2. *Hadar, N., Weintraub, G., Gudes, E., Dolev, S. and Birk, O.S., 2023. GeniePool: genomic database with corresponding annotated samples based on a cloud data lake architecture. Database J. Biol. Databases Curation, 2023, p.baad043.*
3. *Weintraub, G., Hadar, N., Gudes, E., Dolev, S. and Birk, O.S., 2024. GeniePool 2.0: Advancing Variant Analysis Through CHM13- T2T, AlphaMissense, gnomAD V4 Integration, and variant co-occurrence queries. Database J. Biol. Databases Curation. Database, 2024, p.baae130*

This chapter explores the application of the data lake concept in the context of genetic research, specifically focusing on human genetic mutations. We will briefly define a few terms from genetics for readers who may not have a background in biology. Our main goal is to highlight the benefits of our data lake architecture in promoting more efficient and effective research in this field, rather than delving deeply into the specifics of genetic research itself.

## 6.1 Overview and Intuition

A significant amount of genomic data is accessible through services like the NCBI Sequence Read Archive (SRA) [73]. However, this data is stored in a raw format (i.e., FASTQ [74]), making it challenging for geneticists to analyze. Previous efforts have primarily centered on developing frameworks that process raw genomic data in-house and store the aggregated results in a local database [75, 76]. While this approach is reasonable and straightforward, it has several drawbacks:
1. Installing, running, and, most importantly, maintaining such frameworks requires dedicated people with the relevant technical skills not available to many clinical geneticists and researchers.
2. Local databases do not scale well and impose high maintenance costs compared to cloud storage.
3. The in-house approach promotes the appearance of multiple independent repositories (i.e., data silos), whereas the ultimate researchers' goal is to have a single repository with all the relevant data.

To deal with these drawbacks, we have built [GeniePool](GeniePool) – an open-access variant repository. GeniePool is running on the cloud and is accessible to everyone via a simple web interface, so no technical skills are required to use or maintain it (drawback 1). Our repository is stored using cloud data lakes architecture, which provides virtually unlimited scaling while minimizing maintenance costs (drawback 2). Currently, GeniePool is based on more than 60,000 whole-exome samples from SRA, but our highly scalable and inexpensive architecture allows us to import all publicly available human DNA samples and serve as an entry point in genomic analysis for both researchers and clinicians (drawback 3).

From the data management point of view, human genetic data can be represented by 23 arrays (one per chromosome) where every cell in the array can be one of the 4 nucleotide values – G, A, C, T (Table 13). The widely used FASTQ [74] data format consists of nucleotide sequences, similar to those in Table 13.

*Table 13 Human Genetic Data Example*

| chromosome | DNA data |
|---|---|
| 1 | ATGCTTCGGCA… |
| 2 | AGCCCCTCAGG… |
| … | … |

DNA mutations (a.k.a. variants) are defined by comparing the sample DNA data with the *reference* genome data. Common reference genomes are hg19 [77], hg38 [77], and the recently introduced CHM13-T2T [78]. The example of finding a mutation in the given DNA sample is presented in Fig. 22.

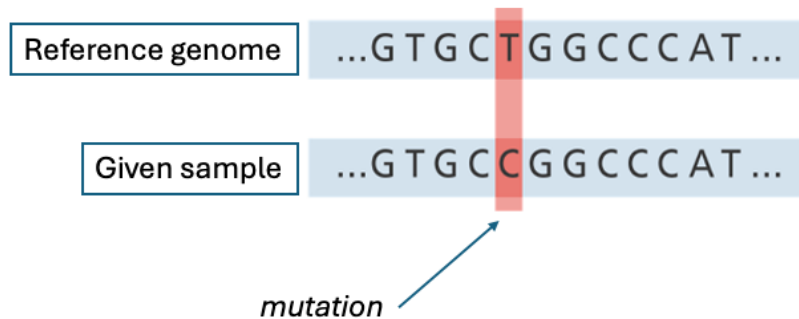

*Figure 22 Genetic Mutation Example*

The challenge we consider in this chapter can be summarized as follows (see Fig. 23):
- Geneticists perform DNA sequencing of their patients in the lab.
- They find many mutations by aligning the patient DNA data with the reference genome (just like in Fig. 22).
- The problem is to understand which of the found mutations are real (pathogenic) mutations and which are benign.
- Our approach is to use publicly available genetic data (e.g., SRA) to confirm or reject the given mutation.
- The idea is to download the open genetic data, pre-process it, and store it in a cloud data lake so we can efficiently answer geneticists' queries.
- The main data management challenge is to be able to find the *tight coverage set* for each given genetic query and thus improve query performance as much as possible.

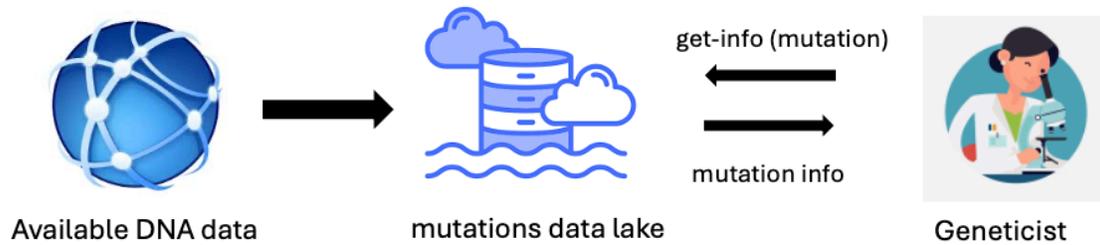
*Figure 23 Genetic Challenge Intuition*

In the next section, we present GeniePool – our approach to the above challenge.

## 6.2 GeniePool

### 6.2.1 Methods

A high-level architecture diagram of GeniePool is presented in Fig. 24, and the example of data transformation in Fig. 25.

The flow contains several important components:
1. SRA stores many human DNA samples uploaded to the repository by genetic researchers as a part of their research. We can represent each entry in SRA as a triple *(id, data, metadata)* where $id \in N, data \in \{'G', 'A', 'C', 'T'\}^*$ and *metadata* is a custom free text.
2. We start by downloading SRA entries and find all the mutations in each sample according to each of the supported reference genomes (hg19, hg38, CHM13-T2T). We use well-established pipelines [80] to convert raw DNA data from FASTQ into VCF format [79]. VCF file can be seen as a relation of the following structure – *(chrom, pos, ref, alt, id)*, where $chrom \in \{1, 2, ..., 22, 'X', 'Y'\}$ is the chromosome, $pos \in N$ is a coordinate within the chromosome, $ref \in \{'G', 'A', 'C', 'T'\}^*$ is the sequence value inside a reference genome at the specified position, $alt \in \{'G', 'A', 'C', 'T'\}^*$ is the sequence value inside the given sample specified position, $id \in N$ is the SRA ID (see Fig. 25).
3. Then, we perform an ETL process that inserts all the found mutations (VCF files from the previous step) into the cloud data lake. Our ETL is implemented with Apache Spark and is running on the AWS EMR platform in "on-demand" mode. The main goal of the ETL is to aggregate mutations such that different SRA IDs will be grouped by their mutations. The result of this step is a relation of type – *(chrom, pos, ref, alt, ids),* where *chrom, pos, ref, alt* are defined as above, and *ids* is a list of SRA IDs. The result is stored as a collection of Parquet files in the AWS S3 cloud data lake. The data flow example of steps 1-3 is presented in Fig. 25.

4. The data in the lake is accessible to the users via REST API and a website (both running on the AWS Elastic Beanstalk platform). User queries (both via API and the website) are of form *get (ref-genome, chrom, coordinate-from, coordinate-to)*. For example, *get (hg38, 7, 117587750, 117587780)*. The result is a JSON object with the mutations found in the given range according to the given reference genome with all the corresponding SRA IDs. The user then can use SRA IDs to access the relevant metadata and assess whether the mutations in question are benign or not.

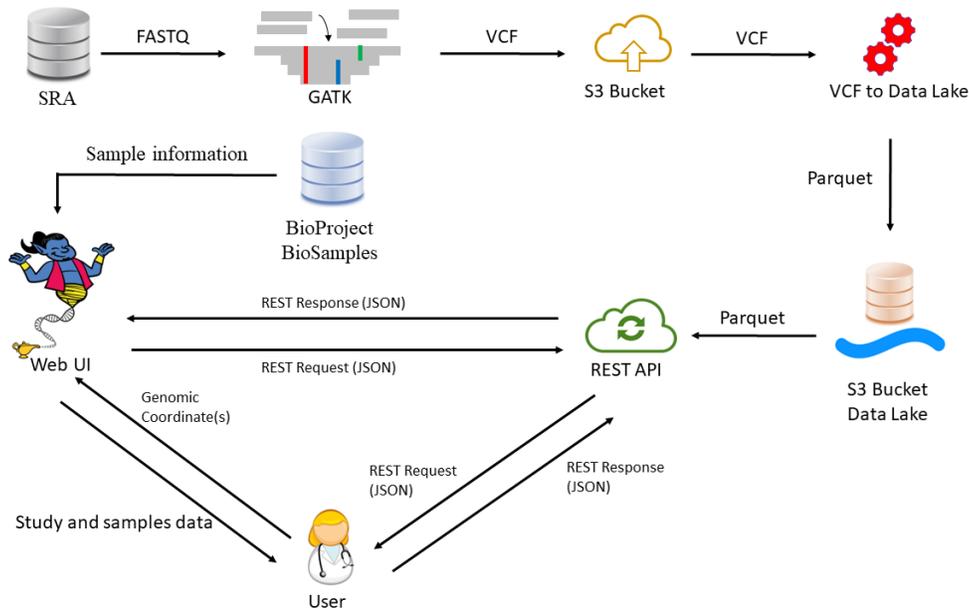

*Figure 24 GeniePool Architecture*

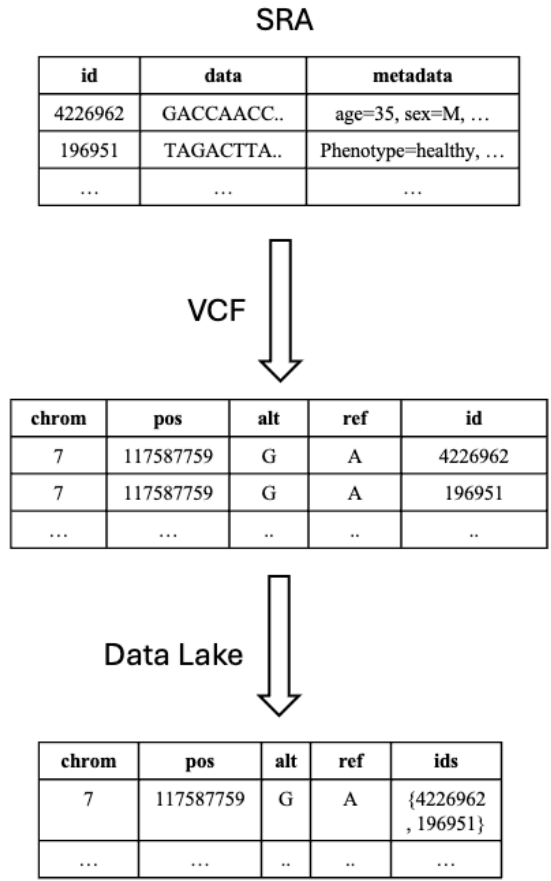

*Figure 25 GeniePool Data Transformation Example*

Now, let us focus again on the main research goal of this thesis: optimizing cloud data lakes' queries by finding the minimal coverage set per query at the minimal cost. In two previous chapters (4, 5), we proposed general-purpose schemes that can be applied to any data lake. In this chapter, our focus is on a particular scenario. Assuming we have a data lake as in Fig. 25 and user queries of type *get (chrom, from, to),* how can we find the best coverage set for each user query?

Our approach is presented in Fig. 26. Our ETL, which reads VCF files and inserts them into the data lake, organizes the data such that we have the data lake partitioned by chromosomes and position buckets. Position buckets are defined as

$$bucket = \left\lceil \frac{pos}{p} \right\rceil \tag{20}$$

where *p* is a constant value defining the maximum number of coordinates per bucket. We tried different values of *p*, and the best results were achieved with *p = 100,000*.

Once the data lake is organized in this way, for each user query *get (chrom, from, to),* we know that the coverage set of the query consists of files located in the given chromosome folder and in the buckets range $[\lfloor\frac{from}{p}\rfloor, \lfloor\frac{to}{p}\rfloor]$. Since our users normally are not interested in ranges larger than *p*, our coverage will contain at most two files, which gives us an almost optimal *tightness degree* ([Definition 4](#)).

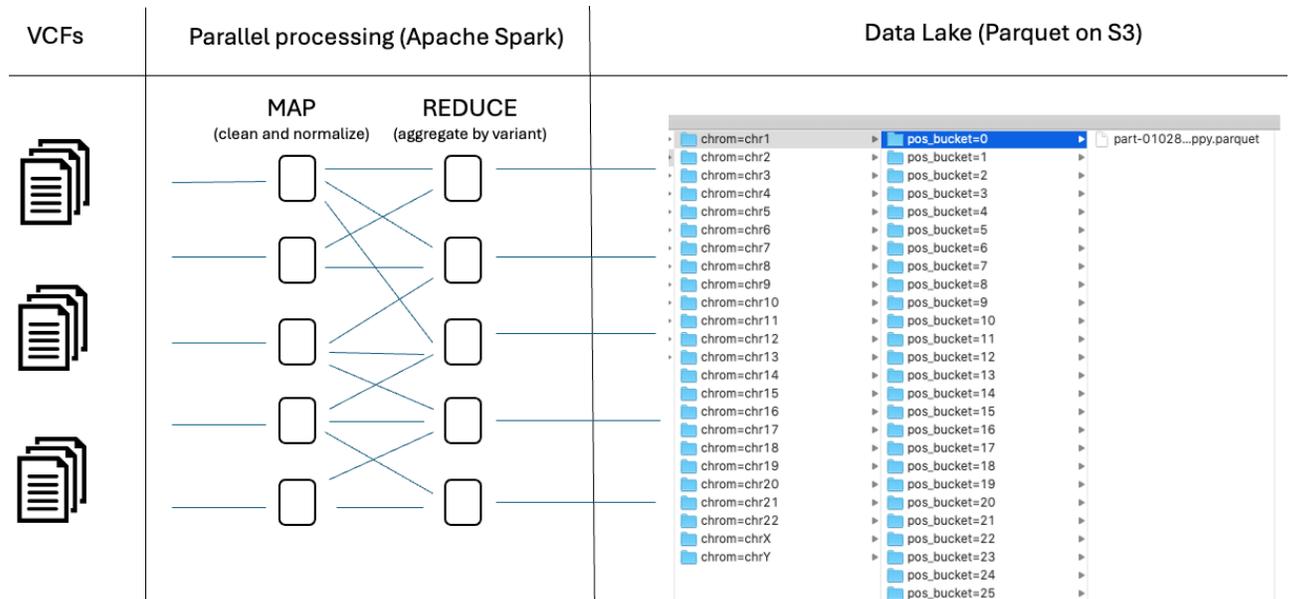

*Figure 26 VCF to Data Lake ETL*

In addition to the core functionality of the GeniePool (finding mutations in SRA samples by given genetic coordinates), we added the following features based on the feedback of our users:
1. dbSNP IDs [81] are attached to the results when available, and search by dbSNP ID is supported.
2. AlphaMissense [82] scores are attached to the results when available, and filtering by the AlphaMissense score is supported.
3. Integration with gnomAD V4 [83] – each variant in the GeniePool response has a direct link to the gnomAD website.
4. Variant co-occurrence – it is possible to find samples where two input variants occurred.

## 6.2.2 Results

### 6.2.2.1 Processed Data

GeniePool currently houses 60,463 samples with almost 3 billion distinct variants. Our data lake consists of 196,613 Parquet files with a total storage size of 411 GB. The data lake is built from the 136,619 gzipped VCF files with a total storage size of 1.6 TB. The ETL job that creates the data lake from the VCF files runs on an AWS EMR cluster of 100 m5.2xlarge nodes with a total running time of around 2 hours.

### 6.2.2.2 Application Interface

Users can query the data by either a web-interface ([Fig. 27](#)) or REST API [84]. In both cases the user needs to specify the reference genome (hg19, hg38, or CHM13-T2T) and genetic coordinates.

*Figure 27 GeniePool Web UI*

### 6.2.2.3 Application Usage

We see that GeniePool attracts many users ([Fig. 28](#)) from many different countries ([Fig. 29](#)). In recent months, there has been an increase in users' activity. The top three countries that are using GeniePool are China, United States, and Israel.

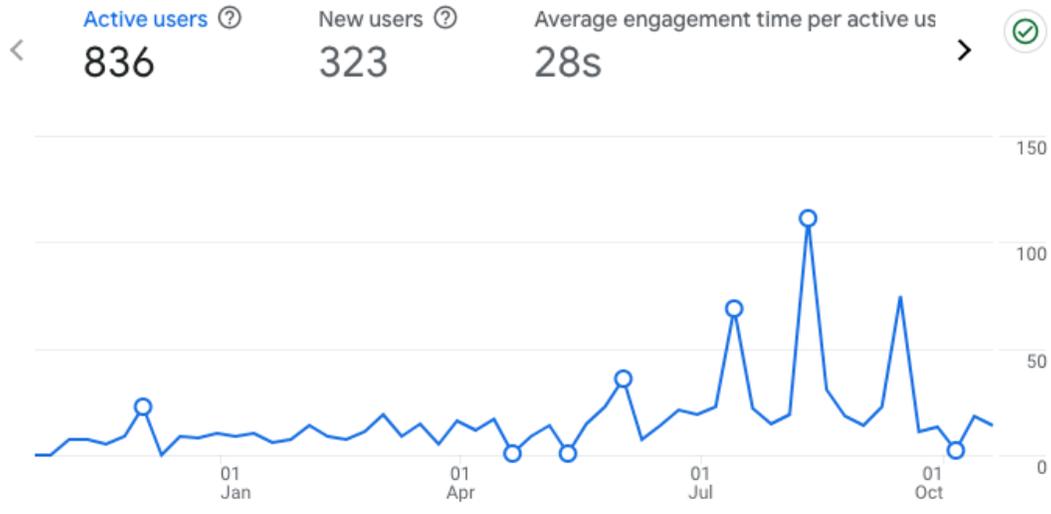

*Figure 28 GeniePool Users Activity Report for October 2023 - October 2024 period (produced by Google Analytics)*

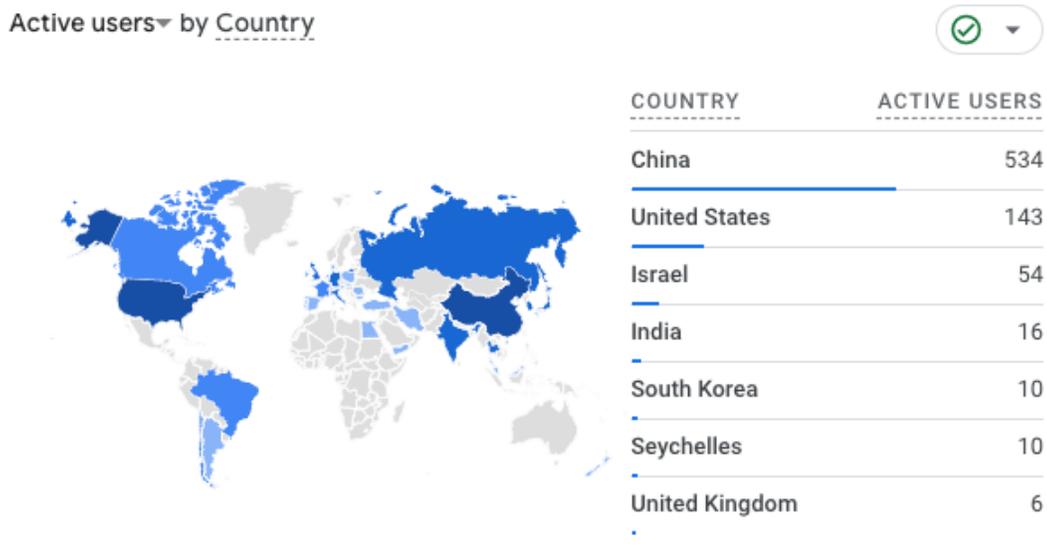

*Figure 29 GeniePool Users by Country report for October 2023 - October 2024 period (produced by Google Analytics)*

6.2.2.4   Application availability

GeniePool source code is available in GitHub repository (https://github.com/geniepool)

GeniePool UI is available at https://geniepool.link/

REST API is available using: http://api.geniepool.link/rest/index/$reference/$coordinates (with hg38/hg19/chm13v2 for reference and chr:start-end for coordinates, e.g. http://api.geniepool.link/rest/index/hg38/1:12345789-123456798).

REST API documentation is available at
https://geniepool.link/GeniePool_API_documentation.pdf

## 6.3   Related Work

The dramatic cost reduction in DNA sequencing in recent years has enabled a lot of opportunities for genetic researchers and clinicians. The main challenge is to classify genetic mutations to be either benign or pathogenic. The prevalent approach is to maintain databases of genomic variants with the corresponding frequency of the variant in a general population [116]. The most popular databases providing such a service are gnomAD [117] and 1000 Genomes [118]. These databases are a great help for geneticists as they allow them to access whether a candidate mutation is present in a population in proportion to the prevalence of the corresponding investigated disease. However, one major problem with these systems is that disease-causing mutations may still appear in them as seemingly non-harmful [119]. That happens, for example, when researchers find a candidate mutation in a few samples in a database, but they cannot know if those samples have a matching phenotype.

This problem is addressed in part by the Sequence Read Archive (SRA) [73], which contains, amongst others, raw sequence data, corresponding phenotypic data, and information regarding the study and the submitters [120]. This data is sufficient for geneticists to access the candidate mutation, but processing and querying it are beyond the capabilities of non-computation-savvy individuals [121].

The idea of preprocess raw sequence data along with the associated metadata has been considered before. For example, in [75, 76, 113], VCF files are preprocessed and stored in a local relational database for further analysis. In [114, 115], a similar approach is used, but the data is stored in a NoSQL database. The main difference between our approach and these systems is that we run on the cloud and require zero effort from the users to start using our service, while in the mentioned systems, users first need to upload their data to the repository. Also, we store genetic data in the cloud data lake rather than DBMS, which results in an order of magnitude lower operational cost [1] while still providing virtually unlimited scalability.

## 6.4 Conclusion

In this chapter, we deal with the problem of query optimization in cloud data lakes in a specific domain – genetic data. Together with genetic researchers, we identified a real need for an application that performs large-scale data processing of genetic data and provides easy access to this data to multiple users.

Our goal was to develop a service that would be scalable, have a low operational cost, and require minimal technical skills from the users. Cloud data lakes perfectly fit these requirements. However, naïve storing of the genetic data in the lake would result in a terrible query performance. We used our theoretical observations from Chapter 2 to design data layout that is optimal for genetic data stored in cloud data lakes. The idea is to organize data in a way such that for any user query, we can detect the tight coverage set of the query without additional cost and read only the file(s) in the coverage set from the lake.

Our service, called GeniePool, is the fruit of a collaboration between computer scientists and geneticists, whose main contributions can be summarized as follows:
1. An open-access variant repository that interactively answers the questions of the type "What kind of variants have been seen in the specific human genomic region? What are corresponding studies and DNA samples?".
2. System usage and maintenance require zero technical skills.
3. The system is built upon cloud data lake architecture, thereby achieving perfect scaling and (extremely) low operational cost.

GeniePool is already used by genetic clinics and research labs in many countries, and we believe that its use will increase as we add more samples to the repository.

Looking forward, the potential of integrating large language models (LLMs) such as ChatGPT [122] for performing complex queries on data stored in our data lake is both exciting and challenging [123]. Despite the transformative impact of LLMs on various fields, their application in genomics is still not mature enough. Our current experiences with LLMs such as ChatGPT4 and Llama 2 [124] have revealed limitations, including inconsistent responses and inaccuracies. However, the evolving capabilities of LLMs promise to revolutionize the way genomic data is queried and analyzed. This could enable sophisticated, intuitive querying mechanisms for genetic researchers, and we expect genomic databases built upon data lake architecture to stand at the forefront of this revolution.

# 7 Conclusions and Future Work

The amount of data being generated is constantly increasing, which has led to the development of new technologies to help manage this growth. One such solution is cloud data lakes, which have proven to be highly successful in dealing with big data challenges. They offer excellent scalability, usability, and cost-effectiveness, but they are not very efficient when it comes to handling queries.

Previous work on query execution performance improvement in cloud data lakes has been mainly on the systems side, and there have not been attempts to provide a solid theoretical foundation for the problem. Thus, one of the main contributions of this thesis is the theoretical framework presented in Section 2.1. We formalize many important definitions related to the cloud data lake architecture, such as query (tight) coverage and tightness and coverage degrees. We use these definitions to formally define the research goal of the thesis as a multi-objective optimization problem where, for each given data lake query, we want to find a coverage set of the query such that the cost of finding this set is minimized, but its tightness degree is maximized. We then approach our research goal by three different techniques in three different domains.

In the "indexing" approach (Chapter 4), our main observation is that if we take any subset of query predicate clauses and intersect their coverage sets, we have a coverage set of the original query (Theorem 1). Then, our problem becomes to find a *balanced* query coverage plan that gives us the optimal subset of clauses. We prove that this "balanced query coverage plan problem" is NP-hard (Theorem 2) but also suggest heuristic algorithms to overcome its hardness. Our coverage plans are based on indexing – we map column values to their files in the data lakes, and that allows us to compute query coverage sets for any data lake query. Indexes are stored in the data lake just like the data and updated and queried with parallel compute engines.

In our future work on indexing, we plan to enhance our system to support index-only accesses [107], such as finding min, max, and counts. Additionally, we aim to support queries across multiple tables by calculating coverage sets through the joining of corresponding index files. An interesting caching technique that we can integrate with our indexing approach is to cache parts of the index on the user's side. This would allow us to execute queries end-to-end, from the client directly to the storage, without needing to access the compute engine. This approach could significantly reduce operational costs.

In the "caching" approach (Chapter 5), we noticed that the query engine at the end of the query processing always knows what the tight coverage of the query was, but this information is not reused in future executions. In addition, many queries in real-world applications tend to have overlaps, so our idea here is to cache tight coverage sets per query and reuse them in subsequent queries contained in one of the cached queries. We show that even if we implement this approach in a trivial way with a linked list, we can achieve tangible improvement in query performance. However, we also develop an enhanced scheme where we map the problem of query containment to a geometric space and use spatial indexes to solve it efficiently.

Our caching scheme is currently limited to conjunctive queries on a single table. In future work, we aim to overcome these limitations. To support more general queries beyond conjunctive ones, we can utilize SMT solvers [112] for query containment checks. For queries involving multiple tables, we will need to design a suitable data structure for our cache, as was discussed in Section 5.2.4. Additionally, we plan to integrate our caching scheme into existing query engines (e.g., [23]) and test it in real-world applications.

Finally, our idea of improving query performance in cloud data lakes by using an optimal query coverage set helps us in the genetic application we designed and developed together with the genetic researchers (Chapter 6). The main idea here is to store genetic data in a cloud data lake in a custom data layout where genetic mutations are stored partitioned by chromosome and genetic coordinates ranges. This way, we can efficiently answer genetic queries by detecting queries' tight coverage sets on the fly.

For our future work in the genetic data domain, we aim to explore the integration of large language models into our system to support complex user queries regarding the genetic data and its metadata stored in the data lake.

In addition to the ideas presented in this thesis, our research has inspired a new direction in the data security domain, where indexes similar to those in our scheme can be securely stored using Bloom Filters [126, 127].

Summarizing, it seems that there is a consensus amongst the researchers that disaggregated architecture is here to stay [22, 96]. In this thesis, we added our contribution to this rapidly developing field. Our ideas span multiple domains, and we believe that they will be useful to the broad research and industry communities.

מטרת המחקר העיקרית של תזה זו היא לגלות שיטות חדשות לאופטימיזציה של שאילתות במערכות אחסון מסוג אגמי נתונים בענן. פיתחנו בהצלחה פתרונות חדשניים בשלושה תחומים שונים: אינדוקס, שמירה במטמון ונתונים גנטיים. פתרונות אלה מבוססים על מסגרת תיאורטית מוצקה ואומתו באמצעות ניסויים מקיפים. בנוסף, הפתרונות לאינדוקס ונתונים גנטיים יושמו בתרחישים בעולם האמיתי.

לצד ההתמקדות המרכזית של המחקר שלנו על טכניקות אופטימיזציה של שאילתות, אנו מאמינים שהתרומות הבאות שלנו עשויות להיות בעלות עניין עצמאי:

- **מודל תיאורטי לשאילתות באגמי נתונים** - ב-[4], הגדרנו מתמטית מספר מאפיינים מרכזיים של אגמי נתונים בענן, כגון כיסוי שאילתות, כיסוי הדוק, דרגות כיסוי ואטימות, ובעיית כיסוי השאילתות המאוזנת. מודל זה מאפשר ניתוח מקיף של ביצועי שאילתות באגמי נתונים בענן ומאפשר לנו להציע פתרונות יעילים. אנו מאמינים שמסגרת תיאורטית זו תהיה בעלת ערך הן לחוקרים והן למהנדסים העובדים עם אגמי נתונים בענן.
- **מיפוי בעיית ההכלה של שאילתות למרחב גיאומטרי** - ב-[6], פיתחנו שיטה לשיפור זמן הריצה של בדיקות מטמון על ידי מיפוי שאילתות לקטעים רב ממדיים. לאחר מכן הפכנו את בעיית ההכלה של קטעים לבעיית חיפוש טווח. ממצא זה עשוי לעניין את קהילת ניהול הנתונים הרחבה.
- **מאגר גנומי** - יצרנו יישום גנטי [7-9] המיישם מושגים תיאורטיים ממחקר זה, ומספק שירות ייחודי לחוקרים גנטיים ולרופאים במאמציהם לטפל בהפרעות גנטיות מורכבות.

התוצאות של תזה זו פורסמו והוצגו במקומות שונים [1-9]. חלק מהפרסומים הללו כוללים "עבודה בתהליך". העבודות המקיפות והעדכניות ביותר עבור כל נושא מחקר הן כדלקמן:

- אינדוקס [4]
- שמירה במטמון [6]
- נתונים גנטיים [8,9]

# תקציר

אגמי נתונים בענן מספקים פתרון מודרני לניהול כמויות גדולות של נתונים. העיקרון הבסיסי מאחורי מערכות אלו הוא הפרדה של שכבות מחשוב ואחסון. בארכיטקטורה זו, אחסון ענן זול מנוצל לאחסון נתונים, בעוד שמנועי מחשוב משמשים לביצוע ניתוח נתונים אלה במצב "על פי דרישה". עם זאת, כדי לבצע חישובים כלשהם על הנתונים, יש להעבירם משכבת האחסון לשכבת המחשוב דרך הרשת עבור כל שאילתה. העברה זו עלולה להשפיע לרעה על ביצועי החישוב ודורשת רוחב פס משמעותי ברשת.

בתזה זו, אנו בוחנים אסטרטגיות שונות לשיפור ביצועי שאילתות בתוך ארכיטקטורת אגמי נתונים בענן. אנו מתחילים בפורמליזציה של הבעיה ומציעים מסגרת תיאורטית פשוטה אך יעילה שמתארת בבירור את הפשרות הנלוות. במרכז המודל שלנו הרעיון של "קבוצת כיסוי של השאילתה", שמוגדר כאוסף הקבצים שצריך לגשת אליהם מהאחסון כדי לענות על שאילתה ספציפית. המטרה שלנו היא לזהות את קבוצת הכיסוי המינימלית לכל שאילתה ולבצע את השאילתה באופן בלעדי על תת-קבוצת קבצים זו. גישה זו מאפשרת לנו לשפר משמעותית את ביצועי השאילתות.

אנו חקרנו עבודות קודמות כדי לזהות פערים בפתרונות קיימים, ולאחר מכן אנו פיתחנו טכניקות חדשות לאופטימיזציה של שאילתות באגמי נתונים בענן בשלושה תחומים שונים:

1. **אינדוקס**: יצירת אינדקס היא שיטה מסורתית לשיפור ביצוע שאילתות במסדי נתונים יחסיים. עם זאת, היא מיושמת רק לעתים נדירות בסביבות ביג דאטה בשל קנה המידה העצום ומורכבות הניהול. אנו פיתחנו שיטת אינדקס חדשנית שתוכננה במיוחד עבור אגמי נתונים בענן, המבוססת על הרעיון של קבוצות כיסוי של שאילתות. הניסיון הראשוני שלנו התמקד בשאילתות מסוג "מחט בערמת שחת" [1, 2]. תכננו אינדקס שממפה את ערכי העמודות לקבוצות הכיסוי של השאילתות שמחפשות את הערכים הללו. אינדקס זה נשמר בתוך האגם הנתונים והוא נוצר ומתעדכן באמצעות אלגוריתמים מקבילים. לצורך הערכה ניסיונית, שיתפנו פעולה עם ארגון גדול המנהל אגמי נתונים בענן בקנה מידה גדול. לאחר מכן, הרחבנו את המודל כדי להתאים לתרחיש כללי [3, 4]. במקרה הכללי, הגדרנו בעיית אופטימיזציה שיכולה להאיץ באופן מוכח את שאילתות באגם הנתונים. הוכחנו שהבעיה הזו היא NP-קשה והצענו גישות היוריסטיות כדי להתגבר על קשיותה. הטמעת אב הטיפוס שלנו, יחד עם הערכות נרחבות המבוססות על TPC-H, הראו שיפור של פי 30 בזמן ביצוע שאילתות.

2. **שמירה במטמון**: מטמון היא שיטה נוספת ידועה לשיפור בביצועי השאילתות, הכוללת בדרך כלל אחסון של תוצאות שאילתות ביניים לשימוש חוזר בשאילתות עוקבות. עם זאת, בסביבת ביג דאטה, גישה זו עלולה להיות בעייתית, שכן תוצאות הביניים עשויות להיות גדולות במיוחד, מה שהופך את האחסון המסורתי לבלתי מעשי. הפתרון שלנו [5, 6] מטפל בבעיה זו על ידי שמירת מטא נתונים (קבוצות כיסוי) במטמון במקום הנתונים בפועל. גישה זו מבוססת על מושגים מתחומי בסיסי הנתונים והגיאומטריה החישובית כאחד. ניסויים המבוססים על TPC-H מוכיחים את יעילות הפתרון שלו.

3. **נתונים גנטיים**: שתי הטכניקות הקודמות מתאימות למטרות כלליות וניתן ליישם אותן על כל מידע יחסי. עם זאת, כאשר אנו יודעים מראש באיזה סוג הנתונים אנו עוסקים, נוכל לתכנן פתרונות יעילים אף יותר. שיתפנו פעולה עם חוקרים מהמעבדה לגנטיקה אנושית באוניברסיטת בן-גוריון כדי ליצור את GeniePool [7-9], מאגר גנומי שנבנה על אגם נתונים בענן. על ידי ארגון אסטרטגי של נתונים גנטיים בענן, אנו יכולים לחשב ביעילות את קבוצת הכיסוי עבור כל שאילתה נתונה. המאגר שהקמנו נמצא בשימוש נרחב על ידי חוקרים ורופאים גנטיים.

# אופטימיזציית שאילתות באגמי נתונים

מחקר לשם מילוי חלקי של הדרישות לקבלת תואר "דוקטור לפילוסופיה"

מאת

גריגורי ויינטראוב

הוגש לסינאט אוניברסיטת בן גוריון בנגב

אישור המנחים פרופ' אהוד גודס ופרופ' שלומי דולב
אישור דיקן בית הספר ללימודי מחקר מתקדמים ע"ש קרייטמן

ל' כסלו תשפ"ה                                             31/12/2024

באר שבע

# אופטימיזציית שאילתות באגמי נתונים

מחקר לשם מילוי חלקי של הדרישות לקבלת תואר "דוקטור לפילוסופיה"

מאת

גריגורי ויינטראוב

הוגש לסינאט אוניברסיטת בן גוריון בנגב

ל' כסלו תשפ"ה       31/12/2024

באר שבע